\documentclass[12pt]{iopart}

\renewcommand{\H} {{\cal H}}

\usepackage[dvips]{graphicx}
\usepackage{color}

\begin{document}
\title{Finite size corrections in the Sherrington-Kirkpatrick model}
\author{T Aspelmeier$^1$, A Billoire$^{2}$, E Marinari$^{3}$, and
M A Moore$^{4}$}
\address{ $^1$Max Planck Institute for Dynamics and Self
Organization, G\"{o}ttingen,
Germany}
\address{$^{2}$Service de physique th\'eorique, CEA Saclay, 91191
Gif-sur-Yvette, France}
\address{$^{3}$Dipartimento di Fisica, INFM and INFN,
Universit\`a di Roma La Sapienza, P. A. Moro 2, 00185 Roma, Italy}
\address{$^{4}$School of Physics and Astronomy, University of Manchester,
Manchester, M13 9Pl, UK}
\begin{abstract}
  We argue that when the number of spins $N$ in the SK model is
  finite, the Parisi scheme can be terminated after $K$
  replica-symmetry breaking steps, where $K(N) \propto N^{1/6}$. We
  have checked this idea by Monte Carlo simulations: we expect the
  typical number of peaks and features $R$ in the (non-bond averaged)
  Parisi overlap function $P_J(q)$ to be of order $2K(N)$, and our
  counting (for samples of size $N$ up to $4096$ spins) gives results
  which are consistent with our arguments. We can estimate the leading
  finite size correction for any thermodynamic quantity by finding its
  $K$ dependence in the Parisi scheme and then replacing $K$ by
  $K(N)$.  Our predictions of how the Edwards-Anderson order parameter
  and the internal energy of the system approach their thermodynamic
  limit compare well with the results of our Monte Carlo simulations.
  The $N$-dependence of the sample-to-sample fluctuations of
  thermodynamic quantities can also be obtained; the total internal
  energy should have sample-to-sample fluctuations of order $N^{1/6}$,
  which is again consistent with the results of our numerical
  simulations.
\end{abstract}

\date{\today}

\section{Introduction}
\label{Intro}

The Sherrington-Kirkpatrick (SK)\cite{SK} model of spin glasses has
been the subject of hundreds of papers. It is the model for which
mean-field theory becomes exact in the thermodynamic limit (i.e. when
$N$, the number of spins in the model, becomes infinite). Parisi's
replica symmetry breaking (RSB) solution\cite{RSB} is now known to be
the correct mean-field solution\cite{Tala}.  Extensive studies, mostly
numerical, have been made of the model at finite $N$
values. Analytically the determination of the properties of the model
at finite $N$ -- the finite-size corrections -- is a much more
challenging task in the low-temperature phase than finding the
mean-field theory.  At finite $N$ all the loop corrections to the
mean-field solution need to be considered.  Because of the massless
modes present in the low-temperature phase\cite{KD} each term in the
loop expansion is infinite. Hence a direct perturbative approach is
impossible. A similar situation applies in finite dimensional spin
glasses, already for the bulk term, when the dimension $d$ is smaller
than six\cite{Moore}, so that one could hope at least that experience
gained in studying finite size effects in the SK model might be
relevant to spin glasses in physical dimensions.

Unfortunately we have been unable to find any systematic theoretical
treatment of the finite-size problem.  However, we have managed to
obtain insights into it by examining the structure of the Parisi
overlap probability distribution function $P_J(q)$ (i.e. {\it the
non-averaged overlap probability distribution function}) at finite $N$
values. $P_J(q)$ is defined as the probability that the overlap of the
spins in two copies of the system with the same realization of the
quenched disorder ${J_{ij}}$ is equal to $q$, i.e.
\begin{equation}
  P_J(q)\equiv 
  \left\langle\delta (q-\frac{1}{N}\sum_{i}\sigma_i \tau_i)
  \right\rangle\;,
  \label{defn}
\end{equation}
where the Hamiltonian of the two-copy system is
\begin{equation}
\H=-\sum_{<ij>}J_{ij}(\sigma_i\sigma_j+\tau_i\tau_j)\;,
\end{equation} 
and the sum runs over all the pairs $ij$ of sites in the system. The
thermal average $\left\langle\cdots\right\rangle $ in Eq. (\ref{defn})
is taken over the Boltzmann weight associated with all the possible
values ($\pm 1$) of the Ising spins $\sigma_i$ and $\tau_i$.  It has
been known for many years that the function $P_J(q)$ is very different
for different realizations of the bonds. In particular it contains a
very variable number $R$ of peaks, humps or shoulders.  In this paper
we shall systematically study the distribution of $R$ and the
dependence of its average on the number of spins $N$.  We shall give
numerical and analytic arguments that the mean number of features $R$
increases as $N^{\mu}$, with $\mu =1/6$ and that $\delta R$, the width
of the distribution of $R$, is $N$ independent for large $N$.
 The next step
of our approach is to argue that the Parisi replica symmetry breaking
scheme, which involves $K$ levels of symmetry breaking (where in order
to achieve a stable solution $K$ has to be taken infinite in the
thermodynamic limit), is stabilized at finite $N$ at a value $K(N)$ by
self-energy contributions (whose $N$ dependence is estimated in
Appendix \ref{appndep}). As a consequence we can estimate $K$ for a
given system size and because $R=2K$ (see Sec. \ref{purestates}), we
can understand the size dependence of the number of peaks/features in
$P_J(q)$.

The next step towards predicting the exponents which give the leading
$N$ dependence of the corrections to the thermodynamic limit of
quantities such as the internal energy per spin $e$ is simply to use
the RSB scheme to compute the dependence of the quantity on $K$.  For
example $e=e_P+O(K^{-4})$, where $e_P$ denotes the value of the
internal energy in the infinite $K$ limit\cite{JK}. Our prescription
for evaluating the exponents of the leading finite size corrections is
to set $K=K(N) \sim N^{1/6}$; this implies that the leading finite
size correction to the thermodynamic limit of the internal energy per
spin should be of order $N^{-2/3}$.  Since arguments of this type do
not have the strength of a theorem and can only be suggestive of the
possible behaviour, we have checked our arguments with extensive Monte
Carlo simulations.  We have computed many quantities at a number of
different values of the temperature. Our results for the internal
energy are reported in Sec.  \ref{energyshift}. That data strongly
supports the value of $2/3$ predicted by our approach for the exponent
of the leading finite-size correction.

Similarly, the Edwards-Anderson order parameter $q_{EA}$ at finite $K$
differs from its infinite $K$ form by a term of order $K^{-2}$.  Hence
we predict that the finite size shift of $q_{EA}$ should be of
$O(1/N^{1/3})$, and we present numerical evidence for this behaviour
in Sec. \ref{qEA}.

Our approach can be used to investigate the sample-to-sample
fluctuations of any quantity by relating them to the sample-to-sample
variation in the number of features in $P_J(q)$, $\delta R$.  For the
internal energy we shall find in Sec. \ref{ssfenergy} numerical
evidence consistent with this approach, together with a discussion of
the behaviour of the sample-to-sample fluctuations in the critical
regime and in the high-temperature phase. Our basic prediction is that
the sample-to-sample fluctuations in the total free energy of a system
of $N$ spins are of order $N^{\Upsilon}$ where the exponent $\Upsilon
=\mu =1/6$. There have been numerous attempts to determine this
exponent, both numerically and analytically, and we review them also
in Sec.\ref{ssfenergy}.

Because the peaks/features in $P_J(q)$ are caused by the overlap of
pure states, in particular those states whose free energies are of
order $k_BT$ from that of the lowest free energy state, one can relate
the number of these pure states to the number of peaks $R$ using the
relation $R=2K$.  This connection is simplified because of the
ultrametric organization of states in the SK model and the details of
the argument are given in Sec. \ref{purestates}.  In
Sec. \ref{finited} we discuss the relation of these ideas with the
behaviour of finite dimensional spin glasses.

\section{Theoretical framework}
\label{TF}

Our Monte Carlo studies of the Parisi overlap probability distribution
function $P_J(q)$ for systems of $N$ spins (with $N$ up to $4096$)
show that the number $R$ of peaks/features is usually quite small, and
that it increases only slowly with $N$, apparently as $R \sim
N^{\mu}$, with $\mu \approx 1/6$.  Our approach to the study of finite
size effects in the SK model is to argue that $R$, the average number
of such peaks/features for a system of size $N$, can be connected to a
truncation of Parisi's RSB scheme at its $K$th step, with $R=2K(N)$.

The Parisi scheme at the $K$-th level of RSB parametrizes the
bond-average of $P_J(q)$, $P(q)$, by a series of delta functions at
various values of $q$, viz $ q_1,q_2, \ldots, q_{K}$;
\begin{equation}
P(q)=\sum_{i=1}^{K}a_i\delta(q-q_i)\;.
\label{PPq}
\end{equation}
The weights of the delta functions $a_i$ and their positions $q_i$ are
the variational parameters that one optimizes to obtain the Parisi
solution.  In the thermodynamic limit, where $N$ goes to infinity, a
Parisi RSB solution with $K>1$ is only stable if $K$ is taken to
infinity. We argue that in finite size systems the self-energy 
corrections to the Parisi solution can stabilize an RSB solution with
a finite value of $K$, and we will argue that $R =2K \sim N^{\mu}$.
(In zero field $P_J(q)=P_J(-q)$ so the number of peaks/features
$R=2K$. However, if $q_1$ just happens to be zero, i.e. there is a
peak at the origin, then $R=2(K-1)+1=2K-1$).

Consider the single-valley replicon correlation function $G_R(i,j)
=\overline{\left\langle S_iS_j\right\rangle _{c}^2}$.  At wavevector
${\bf k}$ its Fourier transform takes the form described in Ref.
\cite{KD}, and at Gaussian order, $G_R(k)=1/k^2$, both for
$T<T_c$ and $T=T_c$.  (Strictly speaking in the SK model the only
possible value which $k$ can take is zero, but we will find it useful
to consider non-zero values of $k$).  Right at $k=0$, $G_R (0)$ is
infinite in the thermodynamic limit.  For finite $N$, the self-energy
corrections neglected at Gaussian order will be shown in Appendix
\ref{appndep} to produce a divergence growing as $N^{1/3}$.
Schematically
\begin{equation}
G_R(k)=\frac{1}{k^2+\Sigma_R}\;,
\label{chi}
\end{equation}
so that the self-energy $\Sigma_R$ is
of order $1/N^{1/3}$.

\begin{figure}
\centering
\includegraphics[width=0.51\textwidth,angle=270]{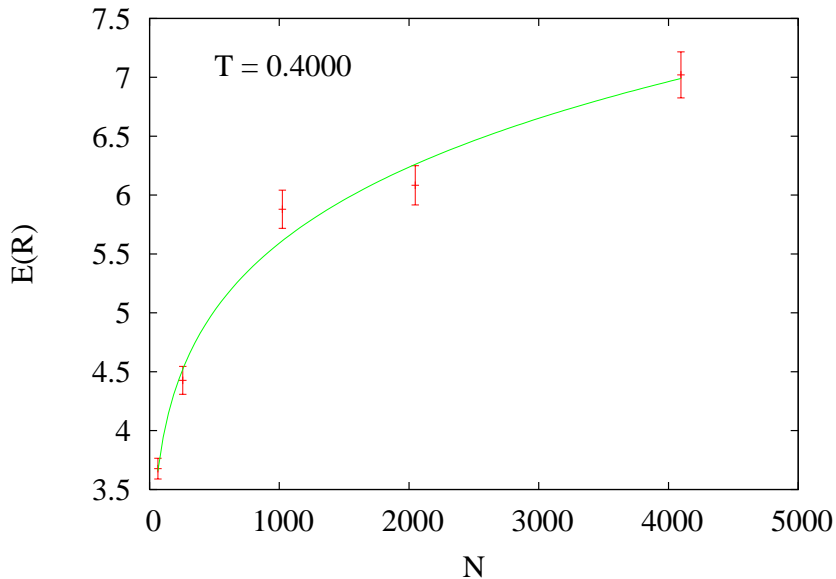}
\caption{Scaling plot of $E(R)$ (the average number of peaks/features
determined by visual inspection of the individual $P_J(q)$) as a
function of $N$, for $T=0.4$. The curve is the best fit to the form
$E(R)=a+b N^c$ with $c=0.17 \pm 0.14$.}
\label{E(R)visual}       
\end{figure}

  Now for finite values of $K$ in the Parisi RSB
scheme, the Gaussian propagator is unstable and behaves as\cite{JK}
\begin{equation}
G_R(k)=\frac{1}{k^2 -\frac{4}{3}\frac{t^2}{(2K+1)^2}}\;,
\label{Instability}
\end{equation}
in the regime near the transition temperature $T_c$ where $t\equiv
1-T/T_c$ is small.  The instability at $k=0$ only disappears when one
takes the infinite $K$ limit.
Our basic idea  is that for finite $N$ this  instability can be removed
by the stabilizing effect of the self-energy $\Sigma_R$. Then if $\Sigma_R
 =c/N^{1/3}$  stability will be achieved when
\begin{equation}
\frac{4}{3}\frac{t^2}{(2K+1)^2} \sim c/N^{1/3}\;.
\label{balance}
\end{equation}
In other words,  when $K =K(N) \sim tN^{1/6}$, there  will be no need
to  break the  symmetry further (at least  to achieve  stability). This
would explain  why the number of peaks/features  in $P_J(q)$ increases
as $N^{1/6}$ (see Fig. \ref{E(R)visual}).

Our procedure to determine the finite size corrections to scaling of
any thermodynamic quantity proceeds in a similar fashion.  First one
obtains from RSB calculations the $K^{\rm th}$ approximation for the
quantity.  Thus the free energy per spin below but near $T_c$ is
 to order $t^5$, and at large values of $K$\;\cite{JK}
\begin{equation}
\Delta f=\left(\frac{1}{6}t^3+\frac{7}{24}t^4+\frac{29}{120}t^5\right)
-\frac{1}{360}t^5\left(\frac{1}{K}\right)^4\;.
\label{Df}
\end{equation}
To estimate the $N$ dependence of the finite size corrections we
replace $K$ by $tN^{1/6}$.  This gives a term in $\Delta f$ which
scales as $t/N^{2/3}$, which is in excellent agreement with numerical
studies\cite{Billoire}.  Just as the self-energy corrections to
Eq. (\ref{Instability}) change the sign of $G_R(0)$, we would expect
that the higher loop corrections to the free energy will also change
the sign of this correction, but not its $N$ dependence.

A similar argument  can be given for other  quantities. The additional
terms in the internal energy per  spin below $T_c$ at order $K$ in the
RSB procedure are\cite{JK}
\begin{equation}
\Delta
u=\left(\frac{1}{2}t^2
+\frac{5}{6}t^3
+\frac{1}{3}t^4\right)
-\frac{1}{72}t^4\left(\frac{1}{K}\right)^4\;.
\label{Du}
\end{equation}
Substituting as before $tN^{1/6}$ for $K$, the finite size corrections
to the internal energy would be expected to be of order $1/N^{2/3}$.

The Edwards-Anderson order parameter\cite{JK} is to
 order $K$
\begin{equation}
q_{EA}=t+t^2-\frac{2}{3(2K+1)^2}t^2\;,
\label{QEA}
\end{equation}
correct to order $t^2$.  It is thus to be expected on substituting for
$K$ that the finite size corrections to $q_{EA}$ are of order
$1/N^{1/3}$. If one defines $q_{EA}$ for finite-size systems as the
value of $q$ at which the Parisi overlap function $P(q)$ peaks, then
such an $N$ dependence is in excellent agreement with both existing
numerical and theoretical arguments\cite{BFM}. We postpone to Section
\ref{qEA} the comparison with the results of our numerical analysis of
the scaling behaviour of $q_{EA}$.

Our approach can be extended to determine the $N$ dependence of
sample-to-sample fluctuations of, say, the internal energy or the free
energy.  In Ref. \cite{AM} it was shown that the variance of the
sample-to-sample fluctuations of the extensive free energy $\delta F$
varies as a quantity $-J(0)$, which at Gaussian order has the property
$-J(k) \approx k^{-2}$.  At finite RSB of order $K$, an exact
expression for this quantity was given in
Ref.~\cite{Francesco}. As shown in Appendix \ref{appJp}, it can
be evaluated when $k^2$ (which is originally a wave vector) is
replaced by the self-energy. One gets that
\begin{equation}
-J \approx N^{1/3}f(t)\;,
\label{var}
\end{equation}
where $f(t)$ is some known function (see Appendix \ref{appJp}).  This
shows that the variance of $\delta F$ is of order $N^{1/3}$, with
typical fluctuations being of order $N^{1/6}$.
  
From this result one can compute the sample-to-sample fluctuations of
$R$.  To do this we shall suppose that the sample-to-sample
fluctuation of $K$ is of order $\delta K$.  Then Eq. (\ref{Df})
implies that the sample-to-sample fluctuation of the extensive free
energy $\delta F$ behaves as
\begin{equation}
\delta F \sim N t^5 \left(\frac{1}{K}\right)^5 \delta K\;.
\label{dK}
\end{equation}
Given that $\delta F \approx N^{1/6}$, it follows that $\delta K$ is
of $O(1)$, that is, independent of $N$.

We have determined the sample-to-sample fluctuations of the internal
energy in the course of our simulations, and their $N$ dependence can
be predicted by extension of these arguments.  From Eq. (\ref{Du}),
the sample-to-sample variation of the full internal energy $\delta U$
is
\begin{equation}
\delta U \approx N t^4 \left(\frac{1}{K}\right)^5 \delta K\;.
\label{dU}
\end{equation}
Substituting for $K$ and $\delta K$, it follows that $\delta U \approx
t^{-1}N^{1/6}$. These sample-to-sample fluctuations appear to diverge
at $T=T_c$, but Eq. \ref{dU} only holds in the RSB region, which is outside
 the critical regime, (which is 
defined
by the limits $N \rightarrow \infty$, $t \rightarrow 0$,
 with $Nt^3$ fixed \cite{Yeo}). 

All  these  arguments  are   intuitive  rather  than  rigorous.  As  a
consequence we  have attempted to check them  by numerical simulations
of the finite size SK model.

\section{The Monte Carlo simulation}
\label{Monte}
We have based our analysis on a large set of numerical data produced
by the large scale parallel tempering simulation of
Ref.~\cite{BiMa1} and \cite{BiMa2}, supplemented by a new
large scale simulation for lattices with $N=2048$ spins.

The quenched random couplings of our system can take the two values
$\pm 1$ with equal probability; the use of such binary couplings
allows to write computer codes that run much faster than, say, when
using quenched random couplings assigned under a Gaussian
distribution.  We assume that the interesting leading scaling behaviour
is the same, for example, when using binary or Gaussian couplings.

We report in Table~\ref{default} the relevant parameters of our
numerical simulations. The temperatures allowed to the parallel
tempering steps are in the range $T\in[0.4, 1.3]$.

\begin{table}[htdp]
\begin{center}
\begin{tabular}{| r | r | r | r |}
\hline $N$ &  $N_{meas}$ & $N_{equi}$ &  $N_J$ \\ 
\hline 64 & 1000 K & 400 K & 1024 \\ 
\hline 128 & 1000 K  & 400 K & 8192 \\  
\hline 256 & 1000 K & 400 K & 1024 \\ 
\hline 512 &  200 K & 200 K & 1024 \\ 
\hline 1024 & 1000 K & 400 K & 1024 \\ 
\hline 2048 & 200 K & 200  K & 512 \\
\hline 4096 & 500 K & 400 K & 256 \\ 
\hline
\end{tabular}
\end{center}
\caption{The relevant parameters of our numerical runs: number of
sites $N$, number of parallel tempering sweeps used for measurements
$N_{meas}$, number of parallel tempering sweeps used for
thermalization $N_{equi}$, and number of disorder samples $N_J$.}
\label{default}
\end{table}

A parallel tempering sweep consists of one Metropolis sweep (all spins
are updated in lexicographic order) followed by a temperature exchange
sweep (we try to exchange adjacent values of $T$ in sequential
order). The balance between the number of sweeps performed for each
disorder sample and the number of disorder samples included has been
chosen cautiously in order to avoid any possible bias due to a non-perfect
 thermalization; we have chosen a safe compromise favouring, at
fixed amount of computer time, the number of sweeps over the number of
disorder samples.  We have checked the quality of thermalization by
monitoring for example the value of $q^2$ as a function of the Monte
Carlo time, starting from an ordered initial spin configurations (all
spins equal to one). For all values of $T$ and $N$ the disorder
averaged data do not drift appreciably already after a couple of
thousands sweeps, i.e. far before we start taking measurements.  A
second important test is provided by the symmetry of the individual
$P_J(q)$'s that is very good for most samples (See
Sect. \ref{Overlap}).

In the rest of this note we will denote by $E(\cdots)$ the average
over the quenched disorder,  $U=Ne$
the total internal energy and and $\delta U=N\Delta$ its standard
deviation, with $N^2\Delta^2\equiv E(U^2)-E(U)^2=N^2
\Bigl(E(e^2)-E(e)^2\Bigr)$.

\section{The structure of $P_J(q)$}
\label{Overlap}
\begin{figure}[htb]
\centering
\includegraphics*[height=4.5cm,angle=270]{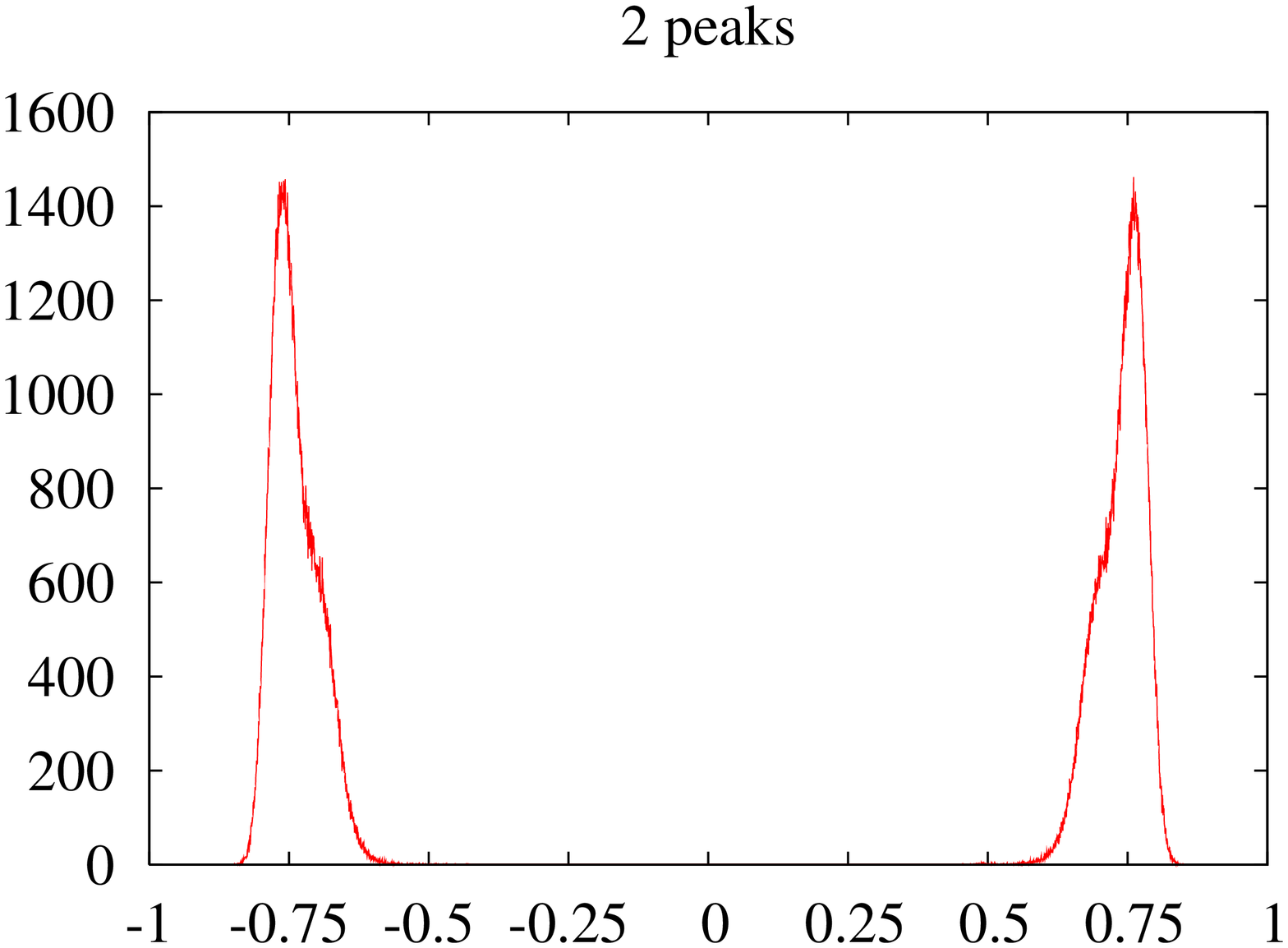}
\hskip                                                              2cm
\includegraphics*[height=4.5cm,angle=270]{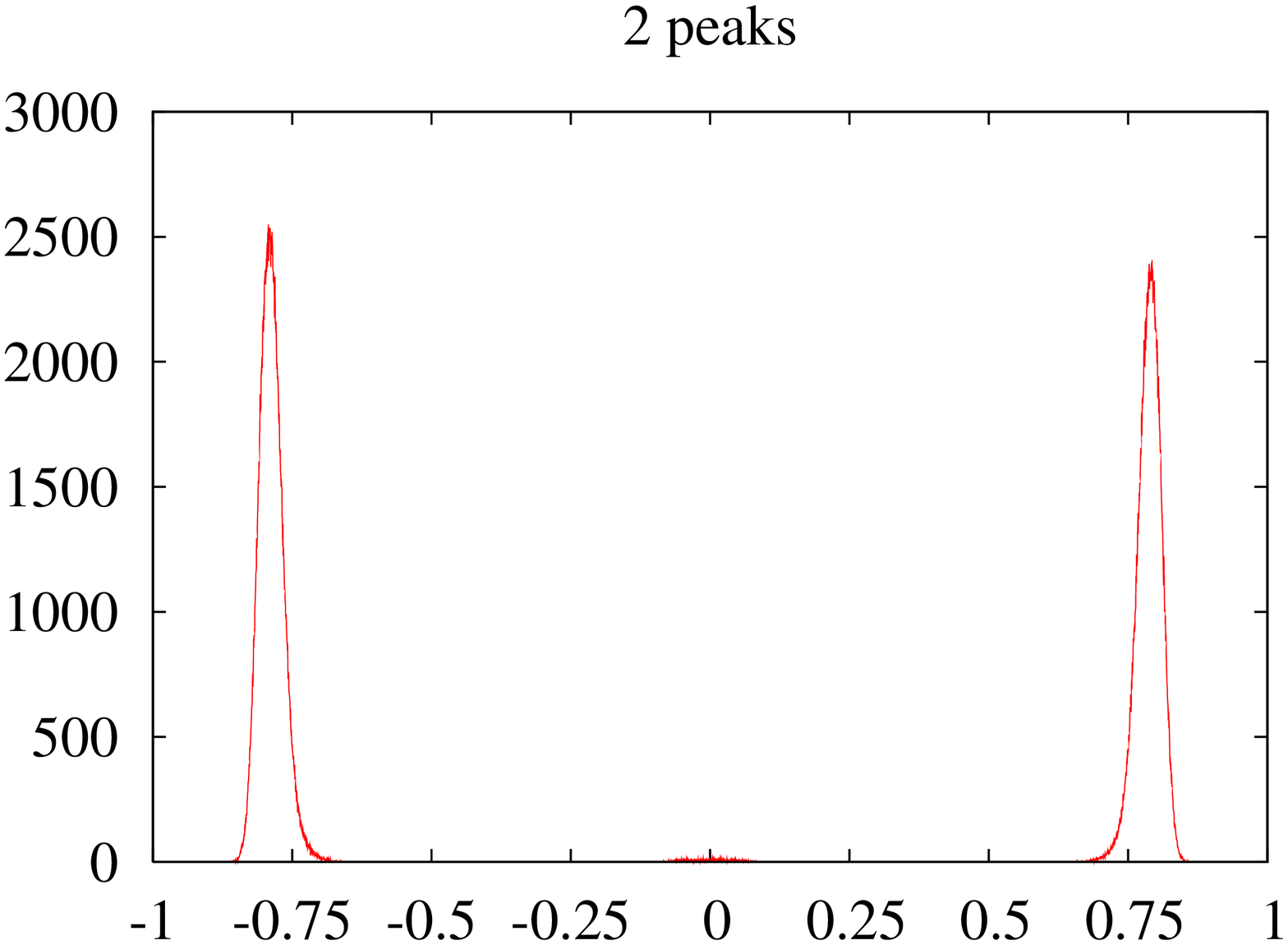}
\hskip -2cm
\vskip                                                              1cm
\includegraphics*[height=4.5cm,angle=270]{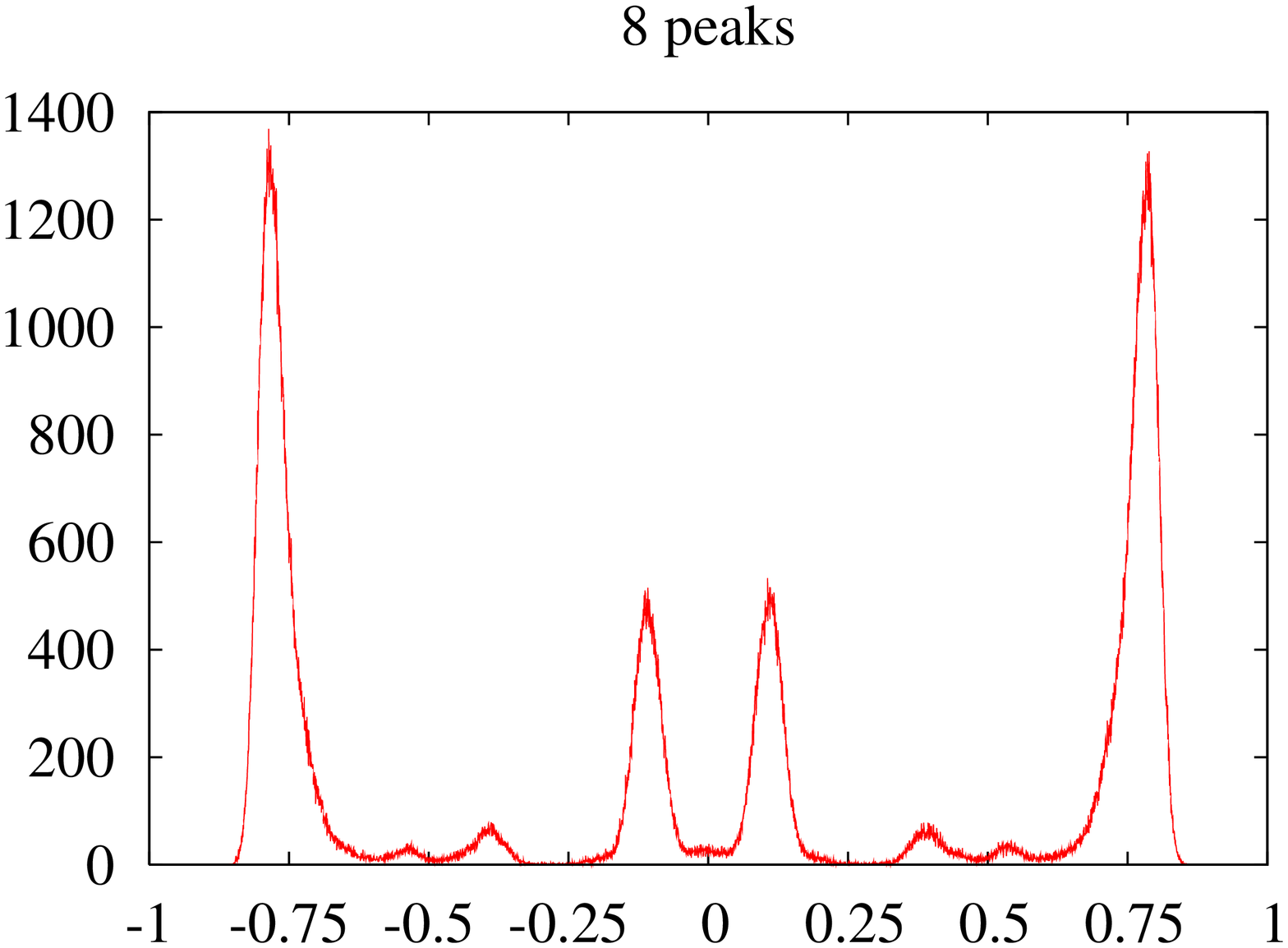}
\hskip                                                              2cm
\includegraphics*[height=4.5cm,angle=270]{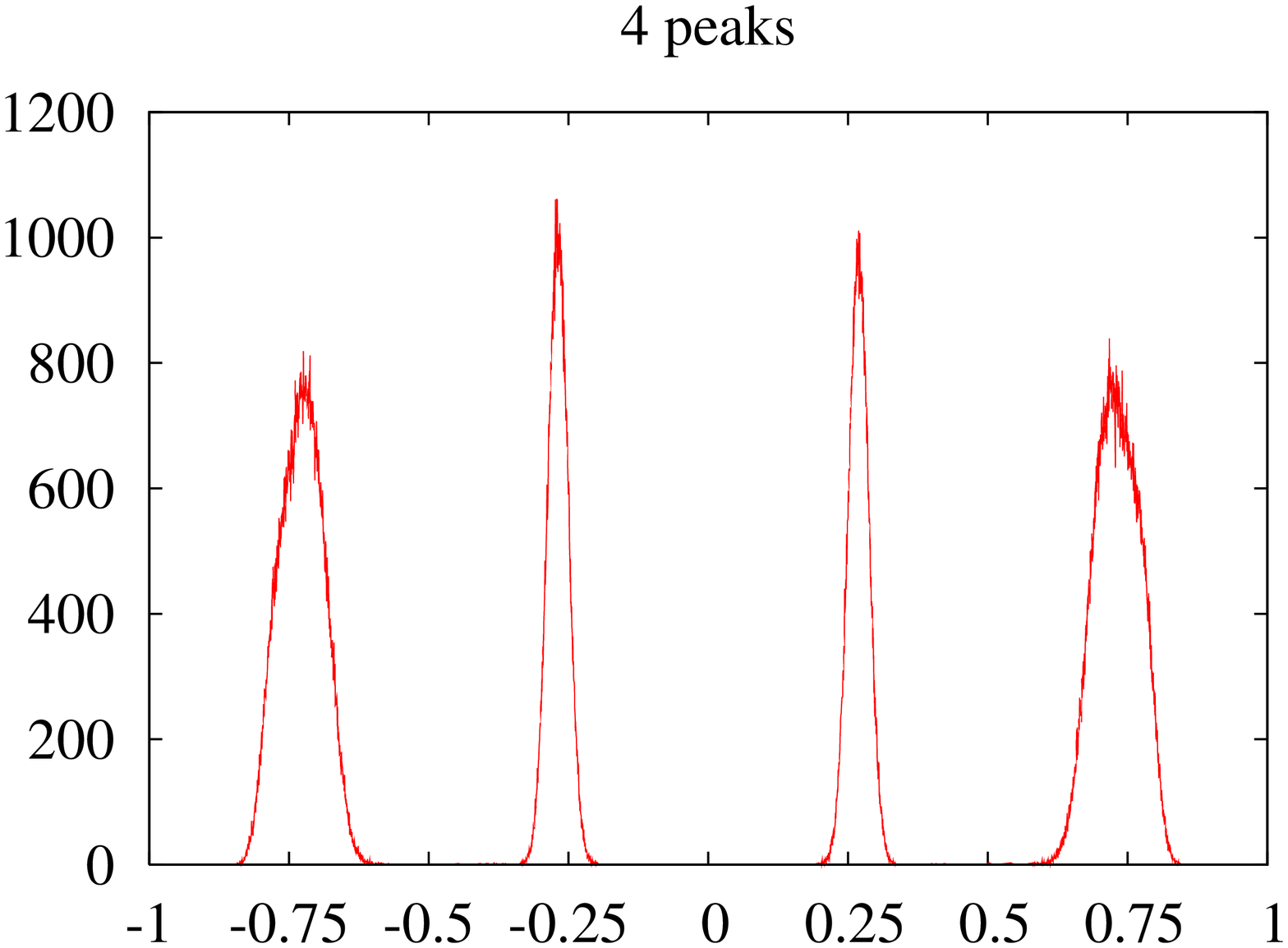}
\hskip -2cm
\vskip                                                              1cm
\includegraphics*[height=4.5cm,angle=270]{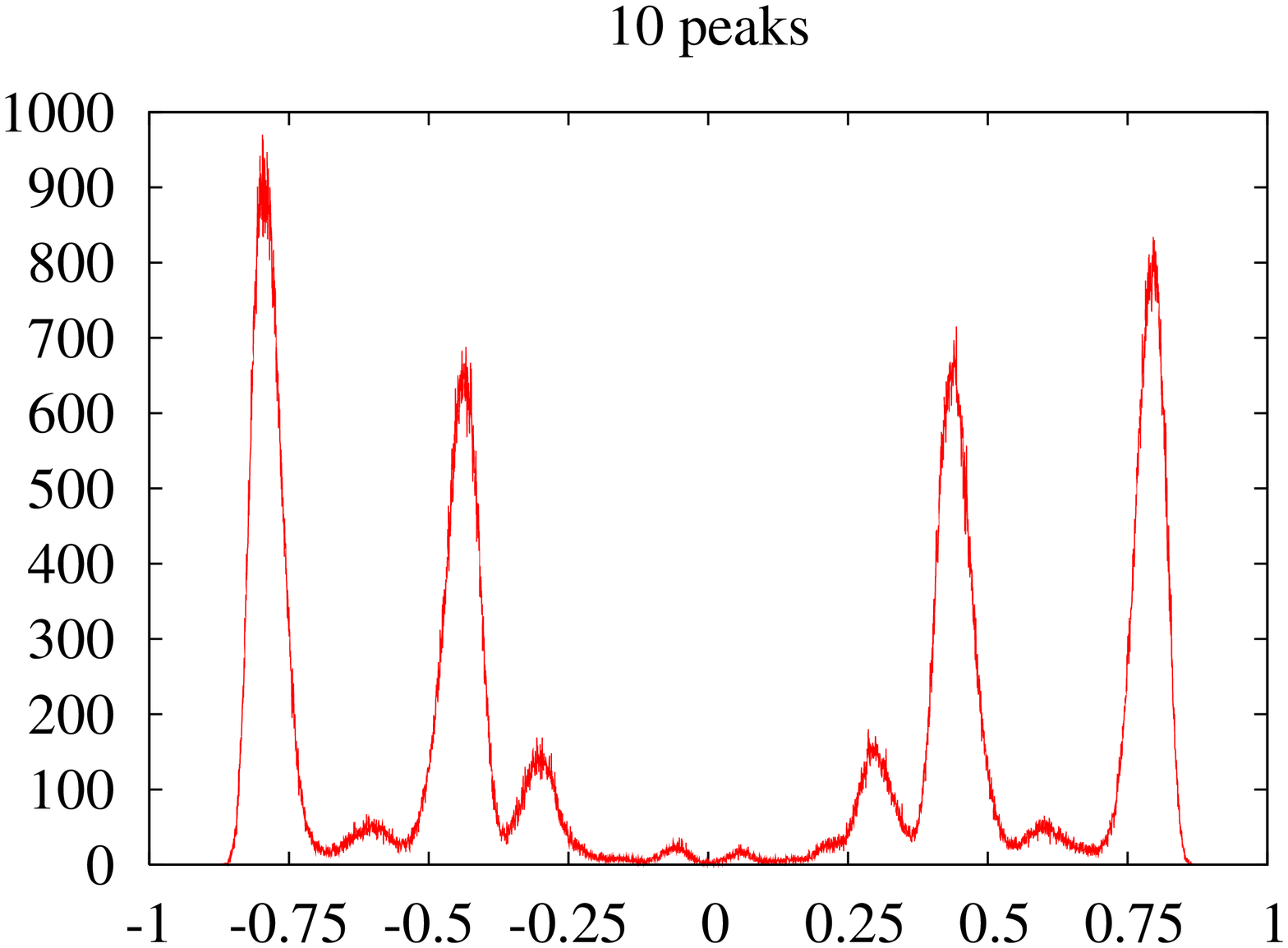}
\hskip                                                              2cm
\includegraphics*[height=4.5cm,angle=270]{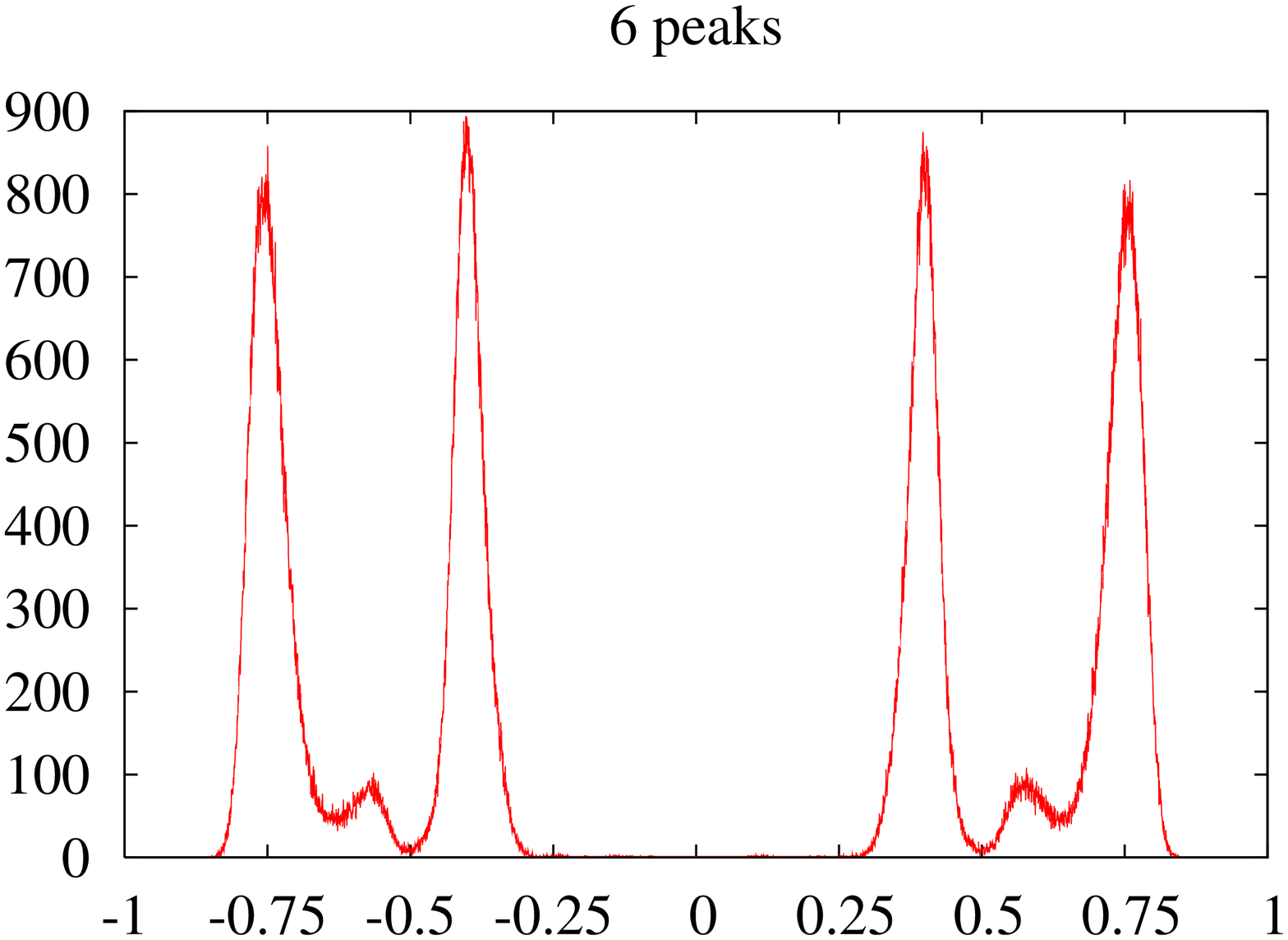}
\hskip -2cm
\vskip                                                              1cm
\includegraphics*[height=4.5cm,angle=270]{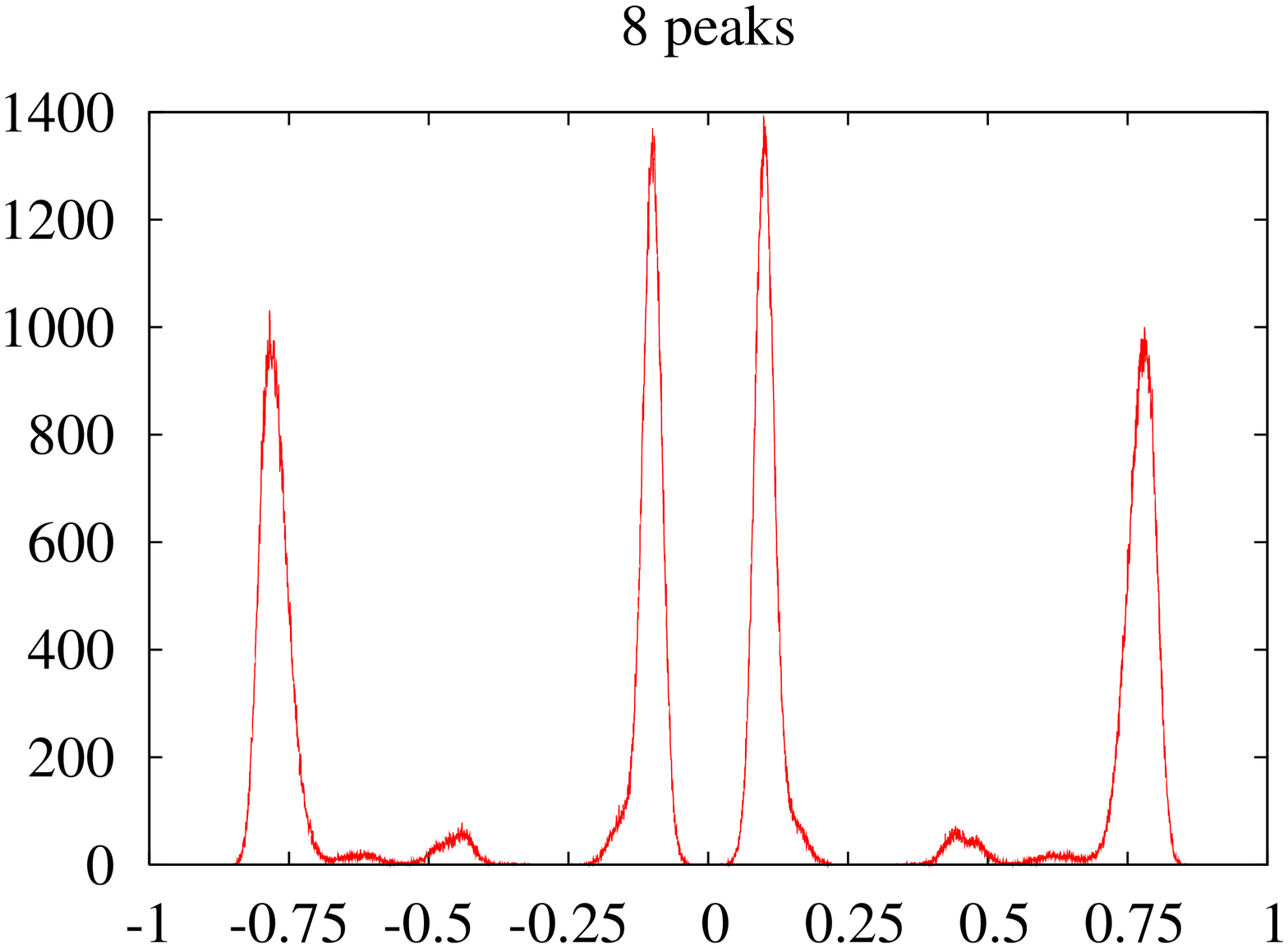}
\hskip                                                              2cm
\includegraphics*[height=4.5cm,angle=270]{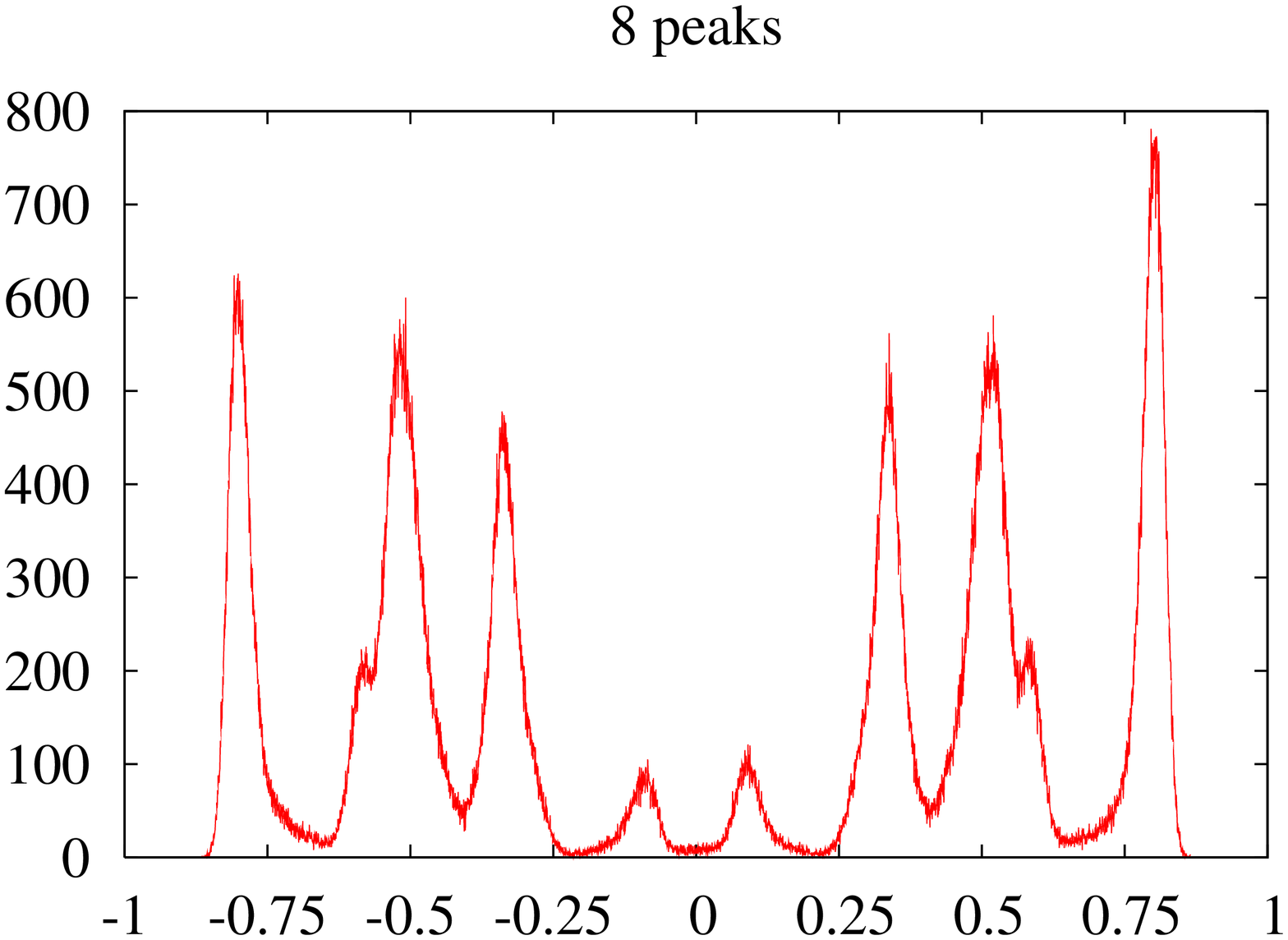}
\caption{$P_J(q)$ for eight different disorder realizations: here
  $N=4096$ and $T=0.4000$.  The symmetry of the plots around $q=0$ is
  a good test of thermalization.  The number of peaks $R$ quoted above
  each figure is the value computed by our computer program.}
\label{P(q)plots}       
\end{figure}

As mentioned before, the function  $P_J(q)$ is very different
for different realizations of the bonds. 
Fig.~\ref{P(q)plots} shows eight
such distributions for the lowest temperature value ($T=0.4$) of the
largest system ($N=4096$) we have simulated. The symmetry of the
plots under inversion of the overlap is excellent: one can see that
even very small peaks appear with their reflected counterpart, and
this is a remarkable check of good thermalization.  The only mild
asymmetries one can see concern the peaks heights, connected to the
population of the different ``pure states to be'', that is a very
difficult quantity to estimate by Monte Carlo integration (just think
about the two peaks in the magnetization distribution for the usual
Ising model in three dimension below the Curie temperature).  Each of
the $P_J$ exhibits a given number $R_J$ of features (well formed
peaks, humps, shoulders on the side of a peak, and so on) that we are
interested to determine in order to compute its disorder expectation
value $E(R)$ and the scaling behavior of $E(R)$ with $N$.

Peaks/feature counting is not at all a trivial issue when dealing with noisy
data. The first assumption must be that the statistical accuracy of
the data set is good enough not to hide important features (in this
respect it is possible that further improvements to the Monte Carlo
scheme could help: one could check for example if using multi-overlap
algorithms\cite{BeJa98,BBJ00} could be of help).  Under this
assumption we have developed a computer code that counts the number of
peaks. The first step is based on smoothing up the (symmetrized) data
for $P_J(q)$.  The second step determines the peaks of the soothed
data: the peaks are defined as local maxima of $P_J(q)$ where a valley
at least lower than a percentage $p$ of the peak height follows the
peak on both sides (we have selected $p=90\%$ and we have relaxed the
depth condition for valleys that include one of the two frontiers of
the support, i.e. $q=\pm 1$).  As a third and last step we impose a
cutoff on the putative peak heights: we discard any peak of height
lower than $1\%$ of the highest peak present in the $P_J(q)$ for the
given disorder sample.  In what follows, we will call ``automatic peak
counting'' this procedure to determine features.

We did not push the coding  to include in our automatic peak
counting the more complex structures which visual inspection spots: an
automatic approach to such a complex task needs great care to
avoid arbitrary choices that could lead to misleading conclusions.

For example it looks clear that the first plot of Fig. \ref{P(q)plots}
(the upper left figure) is characterized by four features, two for
$q>0$ and the two symmetric ones for $q<0$. The first feature is the
clear peak that also the computer code finds, while the second very
clear feature is the shoulder on the peak: this shoulder is naturally
interpreted as a second unresolved peak, too wide to be an isolated
feature, but whose presence is very clear to the observer.  In other
words it is clear that a correct analysis of this feature would lead
to  $R=4$ and that the conclusion $R=2$ reached by our
computer code is not careful enough (as an optimistic remark let us
add that it is possible that the lazy procedure could lead
asymptotically to the same scaling behaviour of a more careful
counting). To be on the safe side, we have carried out the  tedious task of
looking at the first $192$ $P_J(q)$'s for $T=0.4$ and estimating by eye
the number of features $R$ for every graph. This procedure will be
called ``visual inspection'' in what follows.

\begin{figure}
\centering
\includegraphics[width=0.51\textwidth,angle=270]{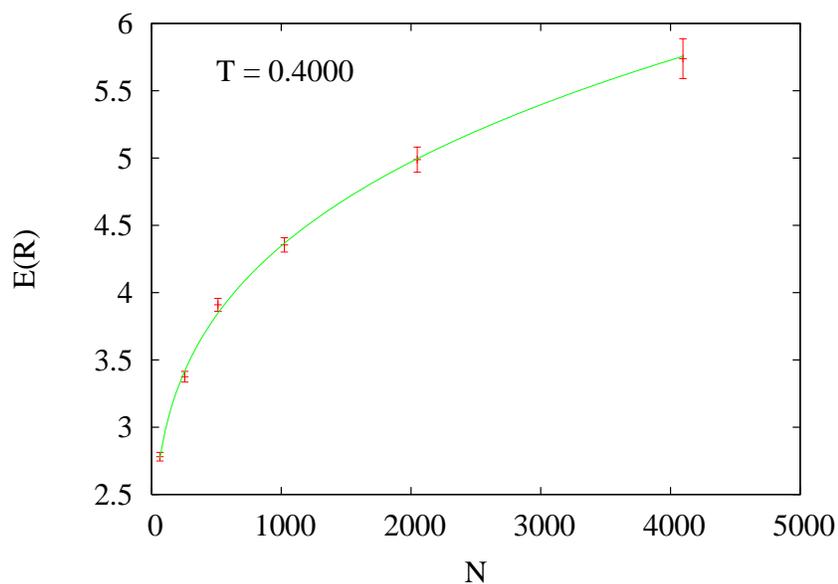}
\caption{Scaling plot of $E(R)$ determined by automatic peak counting,
as a function  of $N$, for $T=0.4$.   The curve is the  fit to the
form $E(R)=a+b N^c$ with $c=0.28 \pm 0.04$.}
\label{E(R)auto}       
\end{figure}

We show the behaviour of $E(R)$ as a function of $N$ for $T=0.4$ in
Fig. \ref{E(R)visual} for the visual inspection and in
Fig. \ref{E(R)auto} for the automatic peak counting.  
The statistical errors are estimated from the fluctuations between the
disorder samples.
Our visual
inspection gives on average between one and two extra features that
are not found by the automatic computer counting.  The best fits to
the form $E(R)=a+b N^c$ for the two cases give the exponents
$c_{visual}=0.17 \pm 0.14$ for the visual inspection and
$c_{automatic}=0.28 \pm 0.04$ for the automatic count.  The difference
between the estimated error bars mainly reflects the number of
disorder samples considered in the two cases. Our prediction
 is $c=1/6\approx 0.17$. Our result for
$c_{visual}$ is on the top of it but it has a huge statistical error,
while our result for
$c_{automatic}$ is not consistent with $1/6$ at more than two standard
deviations (but is of uncertain relevance). It is important
 to notice that here we are dealing with a
quantity that grows very slowly and is very small even for our largest
systems, and that pre-asymptotic effects could be large.

\begin{figure}
\centering
\includegraphics[width=0.51\textwidth,angle=270]{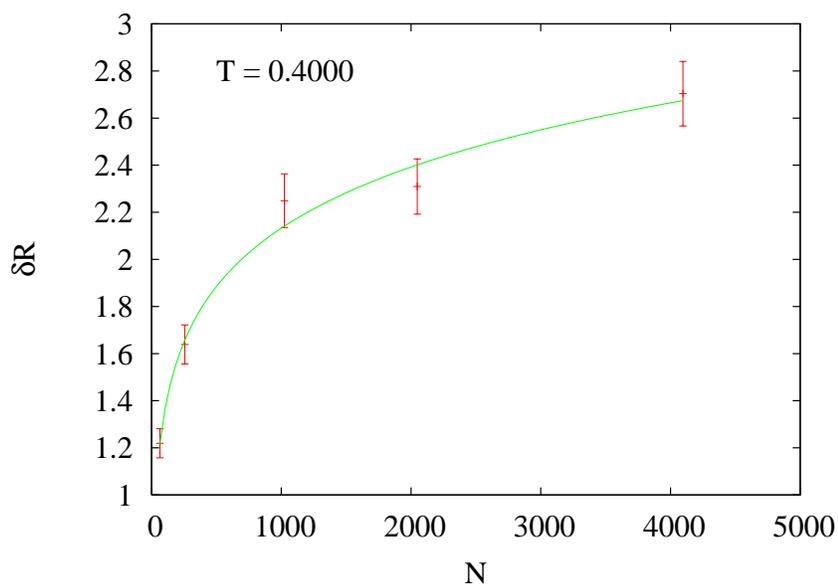}
\caption{Scaling plot of $\delta R$ determined by visual inspection,
as a function of $N$ for $T=0.4$.  The curve drawn is a fit to the
form $\delta R=a+b N^c$.  The best fit is obtained for $c=0.07\pm
0.13$. }
\label{deltaR}       
\end{figure}

In Fig. \ref{deltaR} we  show 
\begin{equation}
  \delta R\equiv [E(R^2)-E(R)^2]^{1/2}\;,
\end{equation} 
i.e.  the fluctuations of $R$ (as determined by visual inspection),
together with the best fit to the form $\delta R = a+b N^c$. The best
fit is obtained for $c=0.07\pm 0.13$: our theoretical estimate
 in Sec. \ref{TF} of the 
value of the exponent $c$ was zero; our numerical
 work is consistent with this estimate, although greater
 precision is really needed before the result can be regarded as definitive.
 
\begin{figure}
\centering
\includegraphics[width=0.51\textwidth,angle=270]{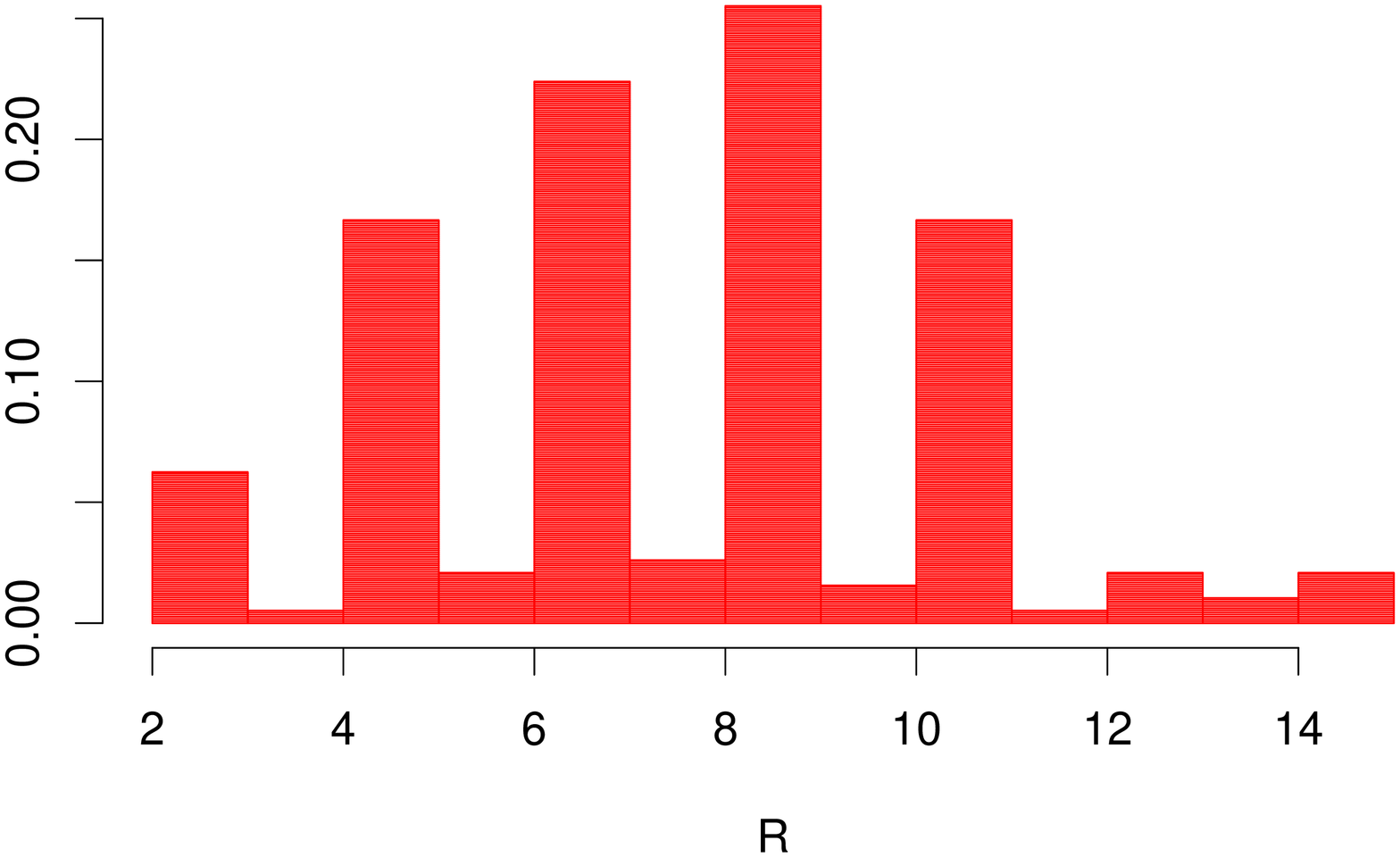}
\caption{Distribution of the number of features $R$ obtained by visual
inspection for $N=4096$ and $T=0.4$. }
\label{peakdist}       
\end{figure}

In Fig. \ref{peakdist} we give the empirical distribution of the number
of features $R$ obtained by visual inspection for $N=4096$ and $T=0.4$:
the distribution is very wide.  It is also clear from the figure that
even values of $R$ are more common than odd values: this is expected,
since odd values are only obtained in cases where $P_J(q)$ has a peak
in $q=0$.

\section{Finite size shift in the energy}
\label{energyshift}
\begin{figure}
\centering
\includegraphics[width=0.51\textwidth,angle=270]{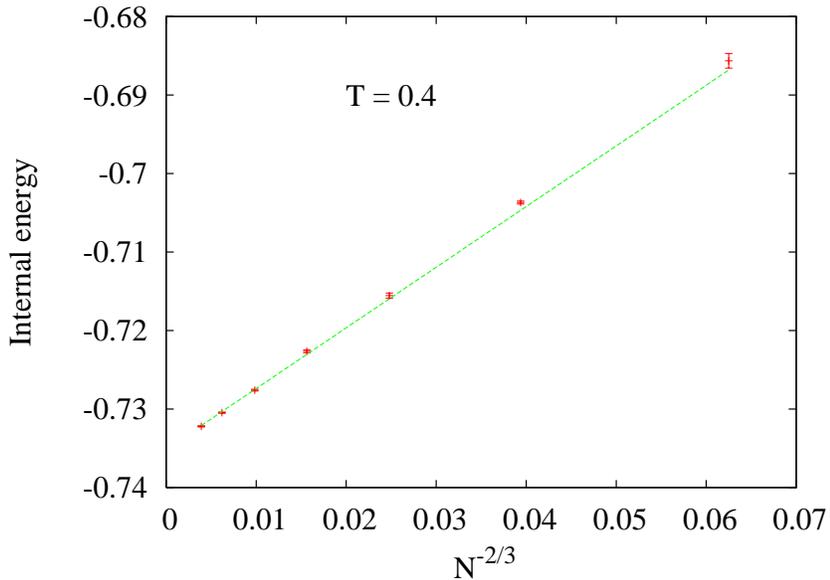}
\caption{The internal energy as a function of $N^{-2/3}$ for $T=0.4$.
The line is a linear fit (using data for $N\geq 256$) as a function of
$N^{-2/3}$ to the data.}
\label{Energyshift}       
\end{figure}

In Fig. \ref{Energyshift} we show the internal energy per spin as a
function of $N^{-2/3}$ for our lowest temperature $T=0.4$ (that we
believe is low enough to be free of effects from the critical point).
The statistical errors are again estimated from the fluctuations between the
disorder samples.

\begin{figure}
\centering
\includegraphics[width=0.51\textwidth,angle=270]{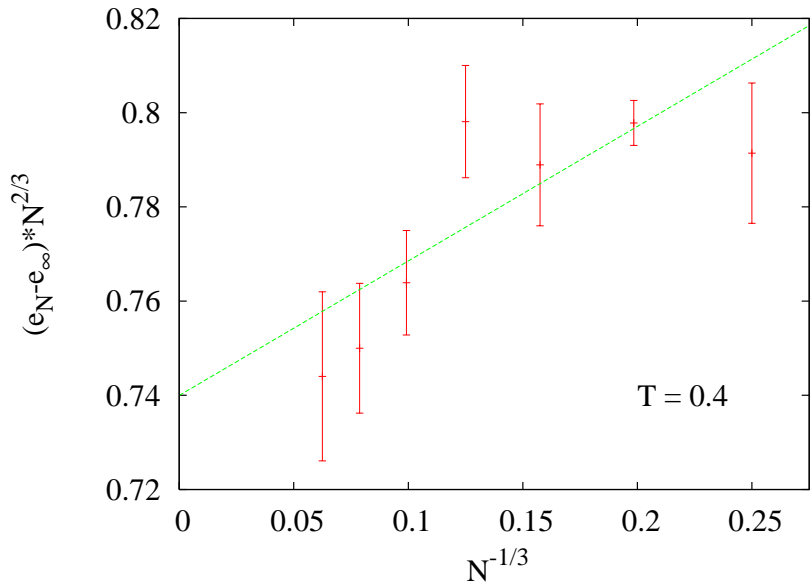}
\caption{ $N^{2/3} ( e_N-e_{\infty})$ as a function of $N^{-1/3}$ for
$T=0.4$. The line is a  linear fit (all data points are included) to
the data, of the form $A+B\;N^{-1/3}$.}
\label{Energycorr}       
\end{figure}

In the same figure we show a fit to the form
$e_N=e_{\infty}+A\;N^{-2/3}$, using the value $e_{\infty}=
-0.735110726$ from Ref. \cite{CrRi}.  The best fit is obtained
for $A =0.77 \pm 0.01$, with a $\chi^2$ of $12$ for $4$ degrees of
freedom.  The presence of slow decaying sub-leading corrections (the
dominant sub-leading contribution behaves like $1/N$, barely faster
than $N^{-2/3}$) explains presumably why the $\chi^2$ is larger than
the number of degrees of freedom.  The importance of the corrections
to the leading behavior (together with some statistical oddness) shows
up in Fig. \ref{Energycorr}, where we plot $N^{2/3} ( e_N-e_{\infty})$
as a function of $1/N^{1/3}$: this is a way to focus on the deviations
from the leading behaviour.  We do believe that the three leftmost odd
looking data points in Fig. \ref{Energycorr} are due to a statistical
fluctuation.  The curve is a fit to  $N^{2/3} (
e_N-e_{\infty}) \propto 1/N^{1/3}$, which is the form we would expect
from a $1/N$ correction to the internal energy per spin.  Based on
Figs. \ref{Energyshift},~\ref{Energycorr} and similar plots at
different temperature values, we conclude that our numerical data are
consistent with an exponent $2/3$ in the whole spin glass phase. We
disagree with the conclusions of Ref. \cite{KaCa} that are based
on data with $36 \leq N \leq 196$, namely a region that is discarded
altogether in our fits (see Ref. \cite{Cecam} for a detailed
comparison).
 
\section{The Edwards-Anderson order parameter $q_{EA}$}
\label{qEA}
We have studied the finite size behaviour of the Edwards--Anderson
order parameter $q_{EA}$, improving the analysis of
Ref. \cite{BFM}.  We define $q_{EA}$ on a finite system as the
location of the maximum of the disorder averaged $P(q)=E(P_J(q))$. The
exact procedure is the following: we first symmetrize our data for
$P(q)$, then we determine the maximum value reached by the function
(we call it $P_{max}$) and finally we compute $q_{EA}$ by means of a
quadratic fit of the data in the range of positive $q$ values such
that $P(q)>0.95 P_{max}$, using the same weights for all data
points. This gives us an estimate of $q_{EA}(N)$ that is not forced to
take discrete values: statistical errors are obtained through a
jackknife analysis (obviously a jackknife approach would not make sense if
$q_{EA}$ was constrained to take discrete values).

\begin{figure}
\centering
\includegraphics[width=0.51\textwidth,angle=270]{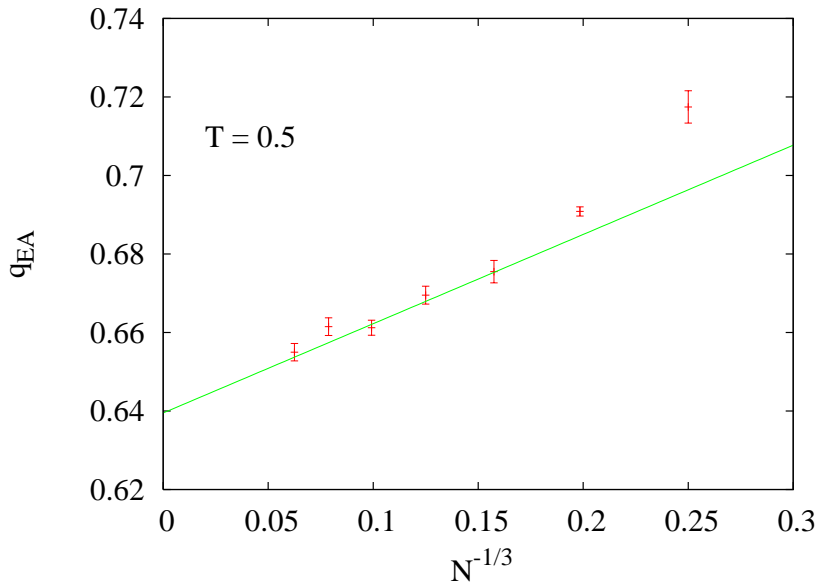}
\caption{The Edwards--Anderson order parameter $q_{EA}(N)$, as defined
  in the text; here $T=0.5$.  The line is for the best fit to the data
  as a linear function of $N^{-1/3}$ (for $N\ge 256$).}
\label{qEAfig}       
\end{figure}

We have compared the values we have obtained for $q_{EA}$ to values
obtained with shorter numerical simulations, and there is an excellent
agreement: this strongly suggests that the procedure we have used to
determine $q_{EA}(N)$ has no appreciable statistical bias.  We show in
Fig. \ref{qEAfig} our data for $q_{EA}(N)$ as a function of $N^{-1/3}$
for $T=0.5$, together with our best fit (using data with $N\ge 256$)
that uses the infinite volume result $q_{EA}= 0.6395$ from
Ref.~\cite{CrRi}.  The value of $\chi^2$ is $1.3$ for $4$
degrees of freedom, suggesting that in this case sub-leading
corrections are not very relevant.  The small $N$ data exhibit larger
corrections from the asymptotic behaviour than the ones for the
internal energy. This is not unexpected since the definition of
$q_{EA}$ on a finite system is involved: for example the intrinsic
resolution of the determination of $q_{EA}$ in a finite system is
$2/N$ (that is close to $0.03$ for $N=64$), a value that is exactly on
the scale of the deviations that we observe.

\section{The sample-to-sample fluctuations of the internal energy}
\label{ssfenergy}
We have also analyzed the fluctuations of the internal energy
between different bond realizations using as a measure
\begin{equation}
\delta  U^2 = N^2      
\Bigl(
E(\left\langle e\right\rangle _J^2)-E(\left\langle e\right\rangle_J)^2
\Bigr)\;.
\end{equation}     
The issue of the scaling behaviour of the free energy fluctuations in
the SK model has
been investigated intensively over the years. Let us define an exponent $\Upsilon$ by
the equation
\begin{equation}
\delta  F^2 = N^2      
\Bigl(
E(\left\langle f\right\rangle _J^2)-E(\left\langle f\right\rangle_J)^2
\Bigr)\sim N^{2\Upsilon}\;.
\end{equation}
Our theoretical approach suggests that $\Upsilon$ should be $\mu
=1/6$.  The first investigation we know of is the numerical work of
Ref. \cite{Cabasino} at $T=0$ which gave (on very small samples)
the result $\Upsilon=0.222$ , compatible with $\Upsilon=1/4$.  The
later theoretical analysis of Ref. \cite{Crea} gives
$\Upsilon=1/6$. However, there is a caveat to this
conclusion. What is effectively calculated is the tail of the
probability distribution of the free energy on the low-free energy
side \cite{RP}; in order to get $\Upsilon$ and one has to assume that
the $N$ dependence of the fluctuations in the tail equals the one of
the standard deviation of the free energy.  Ultimately these analyses
relate the value of $\Upsilon$ to the order of the first non-linear
term in the expansion of the replicated free energy in powers of $n$
(the number of replicas), which is the $n^6$ term \cite{KD}. More
recently, several authors, using exact or heuristic ground states
determination algorithms, have found zero temperature values
compatible\cite{Palassini1,Palassini2,BKM,Boettcher,KKLJH,Pal} with
$\Upsilon=1/4$, some
excluding\cite{Palassini1,Palassini2,Boettcher,KKLJH} the
$\Upsilon=1/6$ value, some not\cite{BKM,Pal}. Analytical arguments in
support of the $\Upsilon=1/4$ value can be found in
Refs. \cite{AMY,BKM}. Finally a recent numerical
simulation\cite{Goethe} using an innovative method obtains
$\Upsilon=2/5$ in the low $T$ phase.  Clearly, the situation is far
from being settled. It should be clear that in the numerical approach
it is extremely hard to distinguish with confidence exponents as close
as $1/4$ and $1/6$ when the range of variations of $N$ is small
(typically one decade), the more so as the functional form of the next
correction is unknown.  A further difficulty is that exact algorithms
are limited to very small systems and heuristic algorithms are
heuristic.

The above results are for the free energy.  The problem for the
internal energy at finite temperature has never, to our knowledge,
been studied numerically; we will consider here finite, low
temperature values, and this will allow us to analyze both the low $T$
phase and the $T\to 0$ limit.

Let $\tilde{e}_J$ be the energy measured at some temperature $T$
during the numerical simulation of a system with a given realization
of the disorder $J$ (this is what we call a sample of our system).
$\tilde{e}_J$ comes from the average of $N_{meas}$ values, and we can
use the quantity
\begin{equation}
\Biggl[   \frac1{N_J}
\sum_J  \tilde  e_J^2-(1/N_J\sum_J  \tilde
e_J)^2 
\Biggr]\;,
\end{equation}
where $N_J$ is the number of samples, to estimate
\begin{equation}
  \Delta^2(T)=E(\left\langle e^2\right\rangle _J)
  - E(\left\langle e\right\rangle _J)^2\;.
\end{equation}
At leading order $\Delta^2(T)$ and $\delta U^2$ are related through
\begin{equation}
  \Delta^2(T) = 
  \frac{\delta U^2}{N^2}
  + \frac{2 \tau_{int}^{(E)} T^2 C_N(T)}{N_{meas} N}\;,
\label{Uno}
\end{equation}
where $\tau_{int}^{(E)}$ is the integrated autocorrelation time for
the energy at temperature $T$, $C_N(T)$ is the specific heat and
$N_{meas}$ is the number of parallel tempering sweeps performed during
the measurement phase of the simulation (we measure the energy after
every Metropolis sweep of the system).

\begin{figure}
\centering
\includegraphics[width=0.51\textwidth,angle=270]{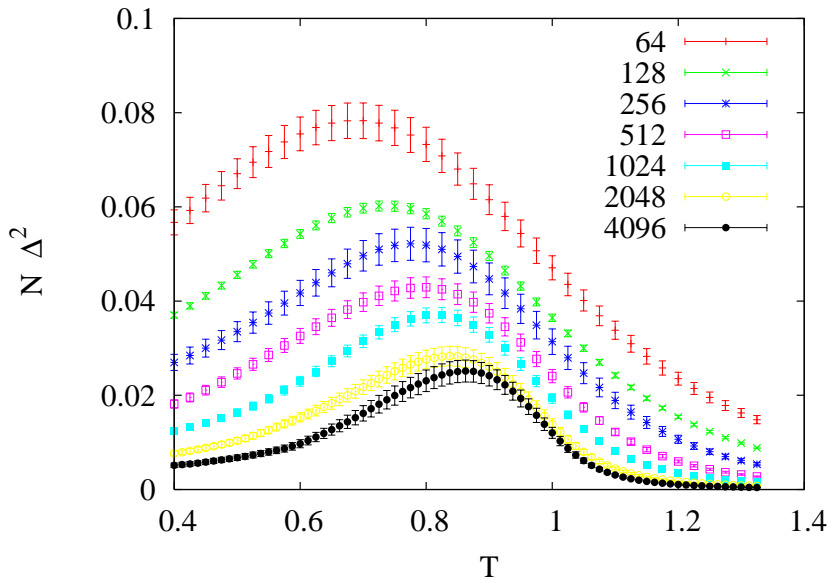}
\caption{The energy fluctuation ${\delta U}^2 /N = N \Delta^2(T)$ as a
  function of $T$, for different values of $N$.}
\label{EF1}       
\end{figure}

It turns out that in our numerical data the second term in
Eq.~\ref{Uno} is negligible, as we have checked by comparing the
estimates of $\Delta^2(T)$ in different numerical simulations for the
same set of disorder couplings (with $N=64, 256$ and $1024$) using the
first $200 K$ parallel tempering (PT) sweeps and the second $1000 K$
PT sweeps after thermalization.  To the best of our knowledge the
autocorrelation time $\tau_{int}^{(E)}$ of the parallel tempering
algorithm has never been measured for the SK model, and our results
show that it is very small.  Our results for $\Delta^2(T)$ as a
function of $T$ can be found in Fig. \ref{EF1}.

\begin{figure}
\centering
\includegraphics[width=0.51\textwidth,angle=270]{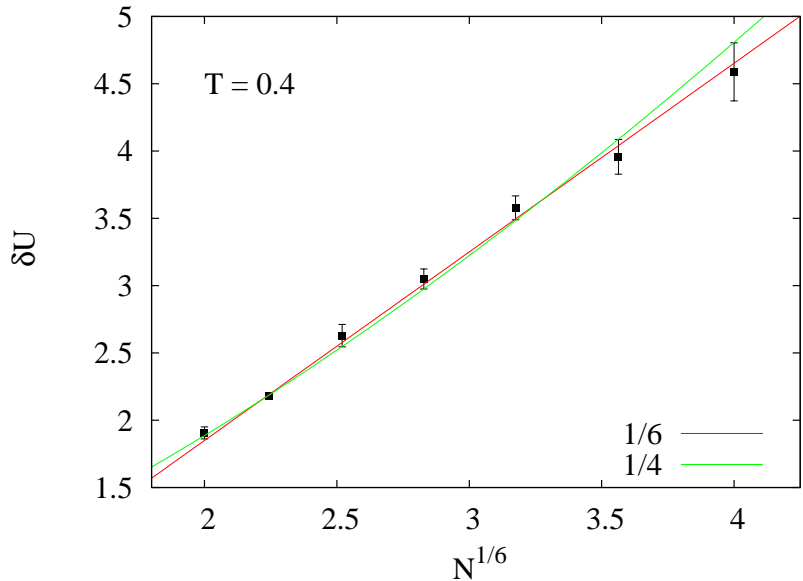}
\caption{The energy fluctuations $\delta U=N \Delta(T)$ as a
  function of $N$, for $T=0.4$.  The two lines are best fits to power
  laws with exponent $1/6$ and $1/4$ respectively.}
\label{EF0.4}       
\end{figure}

Fig. \ref{EF0.4} shows the scaling behaviour of $\delta U$ for our
lowest temperature.  The leading exponent is compatible with the value
$1/6$ but our statistical (and systematic) accuracy is not good
enough to allow us to rule out the value $1/4$: the main culprits 
 are the data points for large  systems (mostly $N=4096$ and $N=2048$), and we
would need a much larger number of samples to have a precise
determination of this exponent from Fig. \ref{EF0.4} only.

\begin{figure}
\centering \includegraphics[width=0.51\textwidth,angle=270]{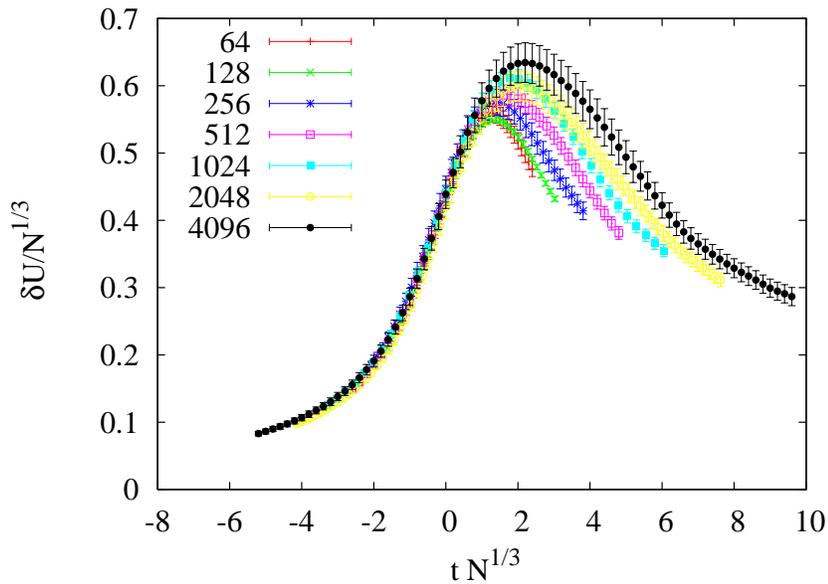}
\caption{Scaling  plot  of $\delta  U/ N^{1/3}$  as  a  function of  
$t N^{1/3}$.}
\label{EFscaling}     
\end{figure}

In the critical region $\delta F \sim f(tN^{1/3})$, where $\delta F$
are the sample to sample fluctuations of the total free
energy\cite{Crea} and $t=1-T/T_c$.  The sample to sample internal
energy fluctuations are related to the $t$ derivative of $\delta F$
and they scale as $N^{1/3} f(tN^{1/3})$, where $f(x)$ goes like $1/x$
at large negative $x$, see Fig.~\ref{EFscaling}, and is a constant at
$x=0$, i.e. $T=T_c$.  The scaling is excellent in the paramagnetic
phase (on the left) and in the spin glass phase where $tN^{1/3}$ is
small, namely before the $\infty$-RSB effects start to be important,
leading to a different behaviour.  Fig. \ref{EFscaling} shows clearly
the two scaling regimes. The $\infty$-RSB effects are only present for
$T<T_c$, in the regime where $Nt^6$ is large \cite{BMTAP} so multiple
pure states can exist. In the finite-size critical regime one has
$Nt^3$ fixed with $t$ going to zero. This makes $Nt^6$ go to zero, so
that in the critical regime RSB effects are absent.

\begin{figure}
\centering \includegraphics[width=0.51\textwidth,angle=270]{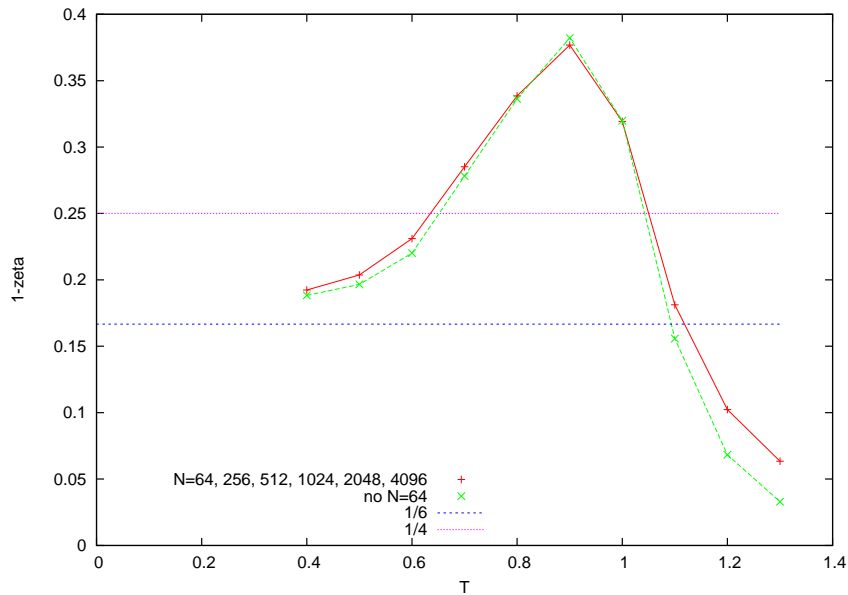}
\caption{Exponent $1-\zeta$ as a function of $T$. We show with the
  continuous lines the results of two best fits, one including all
  vales of $N$ and the other including only values $N\geq 128$.}
\label{EFEnzo}      
\end{figure}

In Fig. \ref{EFEnzo}, we visualize the scaling behaviour of $\delta U$
in a different way; we show the exponent $1-\zeta$ obtained from a fit
of $x=\delta U/(E(e))$ to the form $\propto N^{1-\zeta}$ as a
function of $T$. This plot is consistent with the guess that the
exponent at $T=0$ takes a value of $1/6$: in order to get a value of $1/4$
we should have a very complex $T$ dependence of the effective exponent.  The
situation at $T<T_c$ and exactly at $T=T_c$ is more complicated: it is
possible to see  that finite size effects bring
down the value of the exponent with increasing lattice size, and a
scenario where the exponent is $1/6$ for all $T<T_c$  (but with large 
finite size corrections) is plausible and consistent
with the data.

\section{The number of pure states}
\label{purestates}
It is useful to consider how the peaks in $P_J(q)$ arise.  Suppose we
have just two states, state $a$ and its spin reverse $A$.  Then using
the definition of $P_J(q)$ in Eq.  \ref{defn}, if copy $1$ is in $a$
and copy $2$ is also in $a$ there will be a peak at at $+q_{EA}$. If
both copies are in $A$, there will also be a peak at $+q_{EA}$.
However, if copy $1$ is in state $a$ and copy $2$ is in state $A$,
that will produce a peak at $-q_{EA}$: the same will be true when copy
$1$ is in state $A$ and copy $2$ is in state $a$.

Suppose now we have $4$ states, $a$, $A$, $b$, $B$. Because (in the
infinite volume limit) $q_{EA}$ is the same for all states the
overlaps $aa$, $AA$, $bb$, $BB$ are all $q_{EA}$, and $aA$, $bB$, $Aa$,
$Bb$ are all $-q_{EA}$.  The overlaps $ab$=$q_{12}$=$AB$, and
$aB$=-$q_{12}$=$Ab$, giving 4 peaks in total. (If $q_{12} =0$ we have
$3$ peaks only). However, the peak at $\pm q_{12}$ will not in general
have the same weight as that at $q_{EA}$.

Then, from the above, $2$ states give $2$ peaks and $4$ states give $4$
peaks (if all involved overlaps are large than zero).

With three states $1$, $2$ and $3$ the effects of the ultrametric
organization of states start to play a role.  Besides the peak at
$q_{EA}$, there could be overlaps $q_{12}$, $q_{23}$ and $q_{13}$
making at most $4$ peaks (if all the overlaps are nonzero.). But
ultrametricity says that either all three q's are equal or $2$ are
equal, making at most $3$ peaks ($6$ peaks when one includes the
time-reversed states).

With four states $1$, $2$, $3$ and $4$, besides the peak at $q_{EA}$,
there are the overlaps $q_{12}$, $q_{13}$, $q_{14}$, $q_{23}$,
$q_{24}$, $q_{34}$, but here ultrametricity limits one to $3$ distinct
possibilities, making $4$ peaks in total (and $8$ when one includes the
time-reversed states).

There is clearly a pattern here; if states are organized
ultrametrically their number is equal to the number of peaks of
$P_J(q)$ (plus one if there is a peak at $q =0$). We are thus
predicting that the number of pure states grows with $N$ as $\sim
N^{1/6}$. These pure states are those whose total free energy is of
order $k_BT$ from that of the lowest free energy state. It is only the
overlaps of these low-lying states which can produce detectable
features in $P_J(q)$.

Another way of determining the number of pure states at level $K$ is
to observe that if one starts from the bottom (the leaves) of a
genealogical tree, and moves towards the ancestors,
points that are going up can only meet at the bifurcations of the
tree.  Hence the number of overlaps is equal to the number of levels
$K$ in the tree.

\section{Application to a finite number of dimensions}
\label{finited}
The exponent $\Upsilon$ which determines the sample-to-sample
fluctuations $\delta F$ of the free energy of the SK model has a
significance beyond this model as it appears in the theory of the
interface free energy of {\it finite dimensional} spin glasses.

In Ref.  \cite{AMY} it was shown by going to one-loop order
about the Parisi RSB mean-field solution that the variance of the
interface free energy associated with the change in the free energy on
going from periodic to antiperiodic boundary conditions, $\delta F_{P,
  AP} =F_P -F_{AP}$, is of the form
\begin{equation}
\overline{\delta F^2_{P, AP}}=L^2f(L/M)+\overline{\delta F^2}\;.
\label{freeenergy}
\end{equation}
Here the system is of length $L$ in the $z$ direction, and it is
periodic and of length $M$ in the transverse $d-1$ dimensions.  The
change from periodic to anti-periodic boundary conditions is done by
flipping the sign of the bonds in a hyperplane perpendicular to the
$z$-axis. It follows that $\overline{\delta F_{P, AP}}=0$.
$\overline{\delta F^2}$ is the bond-averaged variance of the free
energy of the SK model containing $N=LM^{d-1}$ spins. Eq.
(\ref{freeenergy}) is valid at least to one loop order, and its form
is probably unchanged whenever the loop expansion is possible; the
loop expansion is well-defined in the low-temperature spin glass state
when $d>6$, but its existence is problematic for $d < 6$ (see Ref. 
\cite{KD}).

The first term in Eq.  (\ref{freeenergy}) is of the standard
aspect-ratio scaling form \cite{Carter, Hartmann}, where the
zero-temperature scaling exponent $\theta$ is equal to $1$.  When $L$
is of order $M$, $\overline{\delta F^2}$ is of order $N^{2\Upsilon}$,
i.e. of order $\sim L^{d/3}$ if $\Upsilon =1/6$.  This term is not of
the standard aspect-ratio scaling form; it depends instead on the
total number of spins in the system.  This reflects the fact that in
RSB situations domain walls have a fractal dimension $d_s$ equal to
$d$, i.e.  they are space filling.  For example in dimension $d=2$,
where we know that we do not have a broken symmetry phase, domain
walls are fractal with $d_s<d$.  The variance of the interface free
energy is {\it dominated} by the SK like term for all $d>6$ provided
that $\Upsilon =1/6$. If $\Upsilon=1/6$, one would expect that
numerical studies of the defect energy in six dimensions would suggest
a value of $\theta$ close to unity; exactly in $d=6$
Boettcher\cite{Boettcher} found that $\theta \approx 1.1\pm 0.1$.
It is possible that when $d<6$ the standard aspect-ratio scaling form
will dominate.  (Of course, when $d<6$, the one-loop expression for
the interface free energy will no longer be adequate). Thus the
dominant term in the $L$ dependence of the interface free energy could
have very different forms above and below six dimensions; note that
$d=6$ plays a special role only if the exponent $\Upsilon$ is exactly
$1/6$.

For $d<6$ the loop expansion about the Parisi RSB state becomes
problematical but it is clearly possible that the essential features
of RSB might survive even in $d<6$, and that the appropriate analytic
approach could allow to make that clear.  Another possibility is that
for $d<6$ the droplet picture of spin glasses \cite{BMreview, FH}
would apply instead, as advocated in Ref. \cite{Moore}, so that
the nature of the spin glass state would change from being RSB like
for $d>6$ to being replica symmetric for $d<6$.  According to this
picture $P_J(q)$ should become just two delta functions at $\pm
q_{EA}$ in the thermodynamic limit, corresponding to just a state and
its time reverse. (In fact, for finite systems in three dimensions,
$P_J(q)$ appears strikingly similar to that of the finite $N$ SK
model\cite{Parisireport}. Unfortunately no systematic study has 
been made as to how the number of peaks, humps and shoulders
evolves with system size; such an investigation could be very
informative as regards the true nature of the three-dimensional spin
glass state.)

We have argued here that for the finite $N$ SK model the replica
symmetry breaking is stabilized at a finite value of $K$ by
self-energy effects. The replica symmetric state of the droplet
picture corresponds to having $K=1$. Thus were the droplet picture to
be the valid description of spin glasses below six dimensions (and
some of the authors of this paper would argue that this is unlikely!),
the same mechanism could stabilize the replica symmetric state. While
perturbatively there seems to be no way that the replica symmetric
state could be stable, it is possible that if the full self-energy
corrections about that state could be included into the calculation,
then replica symmetry might be maintained \cite{Moore}.

\section*{Acknowledgments}
We thank Giorgio Parisi for interesting verbal and virtual
conversations about many of the topics discussed in this note.  We
thank Marc M\'ezard for useful comments and advice, and Jean-Philippe
Bouchaud for discussions. We also thank Andrea Crisanti and Tommaso
Rizzo for allowing us to analyze their numerical data.

\appendix
\section{The $N$ dependence of the self-energy}
\label{appndep}
In this Appendix we shall estimate the $N$ dependence of the
self-energy $\Sigma_R$. This is needed for our argument that the
number of features $R$ scale as $N^{1/6}$.  A direct calculation of
$\Sigma_R$ would be impractical: it would involve summing diagrams to
all orders.  Because of that we will obtain the $N$ dependence of
$\Sigma_R$ indirectly via a study of the TAP equations \cite{TAP} of
the model.

The TAP equations provide a non-replica way  of finding single-valley
correlations. For the magnetization $m_i$ at site $i$ within
a single state they give
\begin{equation}
m_i=\tanh  \left(  \beta  \sum_{j}  J_{ij}m_j-\beta  m_i\sum_jJ_{ij}^2
(1-m_j^2)+\beta h_i\right)\;.
\label{TAP}
\end{equation}
The spin glass susceptibility is defined as
\begin{equation}
 \chi_{\rm SG} \equiv  \frac{1}{N}  \sum_{i,j}\left(\frac{\partial m_i}{\partial
 \beta h_j}\right)^2\;.
\label{corr}
\end{equation}
When the right-hand side is bond-averaged over the exchange
interactions $J_{ij}$ we obtain $G_R(0)$.
In the following we will
only consider the case of zero magnetic field, $h_i=0$.

It is convenient to express $\chi_{\rm SG}$ in terms of the eigenvalues of
the Hessian matrix of the second derivatives of the TAP free
energy\cite{BM0}:
\begin{eqnarray}
A_{ij}&\equiv&\partial^2(\beta     F_{TAP})/\partial     m_i\partial    m_j
=-2\beta^2J_{ij}^2m_im_j                 -\beta                 J_{ij}
\nonumber\\&+&\left(\beta^2\sum_k
J_{ik}^2((1-m_k^2)+(1-m_i^2)^{-1}\right) \delta_{ij}\;.
\label{Hessian}
\end{eqnarray}
In terms of the eigenvalues $\lambda$ of ${\bf A}$,
\begin{equation} 
\chi_{\rm SG}=\frac 1N \sum_{\lambda}\frac{1}{\lambda^2}\;,
\label{chiSG}
\end{equation}
which in terms of the density of states $\rho (\lambda)$ becomes
\begin{equation}
\chi_{\rm SG}=\lim_{N  \to  \infty}\int_{\lambda_{min}}^{\infty}d\lambda  \,
\rho(\lambda)/\lambda^2\;.
\label{eig}
\end{equation}
The first  term on the  right hand side  of Eq. (\ref{Hessian})  is of
order $1/N$  and is smaller than  the other terms which  are either of
order  $1/N^{1/2}$,  or on  the  diagonal, of  order  1.   It will  be
dropped. (We are focusing in this work on the low-lying TAP states --
the pure states  -- where the mechanism for  splitting off an isolated
eigenvalue as  in Ref.~\cite{ABM} cannot operate.)   A stable solution
of the TAP equations which  corresponds to a minimum requires all
the  eigenvalues of the  matrix ${\bf  A}$ to  be positive.   For pure
states,  $\rho(\lambda)$ is  non-zero right  to the  origin;  at small
$\lambda$ (see Ref. \cite{BM0}) one has that
\begin{equation}
\rho(\lambda)=\frac{1}{\pi}\left(\frac{T}{T_c}\right)^3
[\frac{1}{N}\sum_i\left(1-m_i^2\right)^{3}]^{-1/2}\lambda^{\frac{1}{2}}\;.
\label{smalllambda}
\end{equation}
With this  form for $\rho(\lambda)$,  the integral in  Eq. (\ref{eig})
would be divergent without its  lower cutoff at $\lambda_{min}$. The $N$
dependence of $\lambda_{min} $ itself can be estimated by setting
\begin{equation}
1=N\int^{\lambda_{min}}_{0}d\lambda \, \rho(\lambda)\;,
\label{lowerlimit}
\end{equation}
which means  that $\lambda_{min}  \sim N^{-2/3}$. Using this  result, we
can  estimate the $N$  dependence of  $\chi_{\rm SG}$ as  $N^{1/3}$ using
Eq. (\ref{eig}). Notice that this result would also apply at $T_c$.

In fact there is a very simple direct argument for the behaviour at
$T_c$.  According to Refs. \cite{Yeo, Moore} for $T>T_c$,
\begin{equation}
  \chi_{\rm SG}=\frac{1}{|t|}f(N|t|^3)\;,
\label{Tc}
\end{equation}
so that as $|t|$ goes to zero, that is, at $T_c$, $\chi_{\rm SG} \sim N^{1/3}$.

We would not expect that bond-averaging $\chi_{\rm SG}$ to get $G_R(0)$
will modify this $N$ dependence since single-valley quantities are
expected to be self-averaging.

Thus for $T \le T_c$, the single valley spin glass susceptibility
$G_R(0)$ diverges as $N^{1/3}$.  This implies that the typical value
of $K$, the order of replica symmetry breaking in a finite system of
$N$ spins, will be via Eq.  (\ref{balance}) of order $tN^{1/6}$.  The
data in Fig.  \ref{E(R)visual} is clearly consistent with this
expectation.

\begin{figure}
\includegraphics[width=0.71\textwidth]{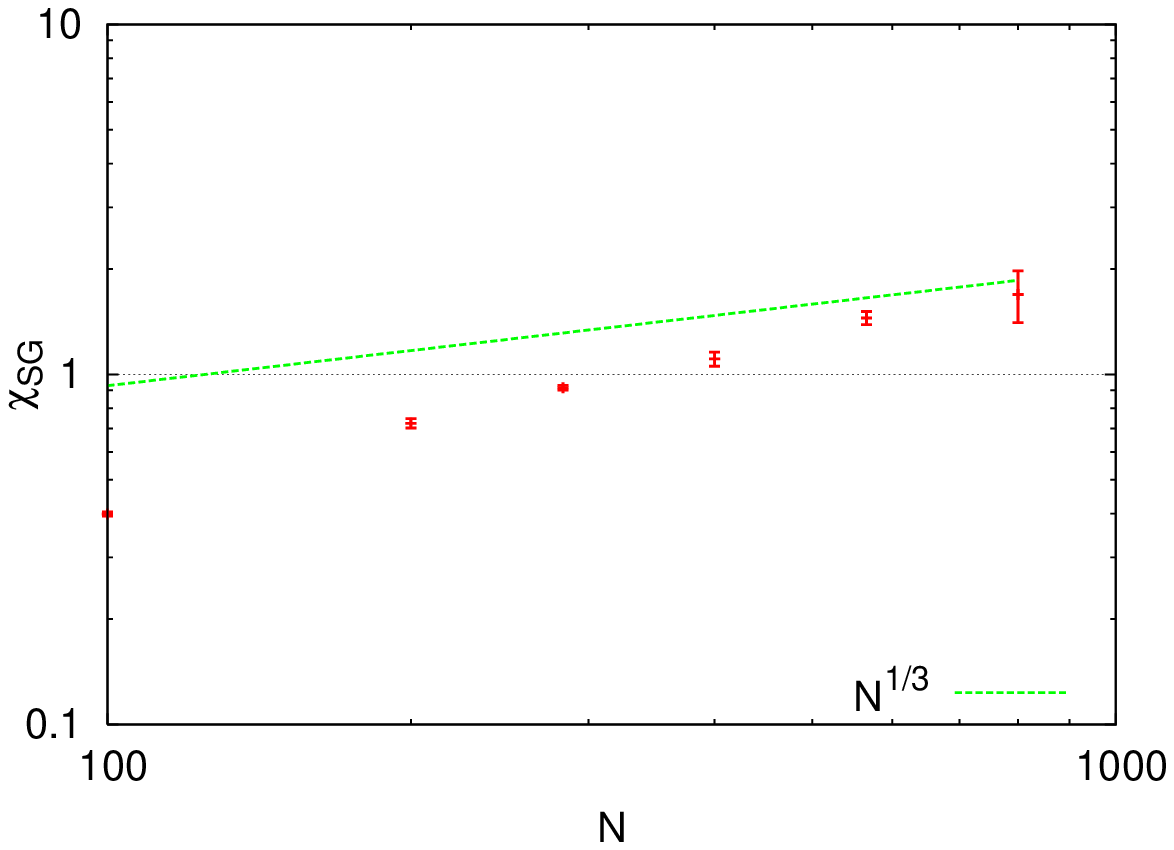}
\caption{The spin glass susceptibility $\chi_{\rm SG}$ as a
function of the system size appears to grow faster than the predicted
$N^{1/3}$. This is due to finite size effects, see text.}
\label{chiSGfig}
\end{figure}

\begin{figure}
\includegraphics[width=0.71\textwidth]{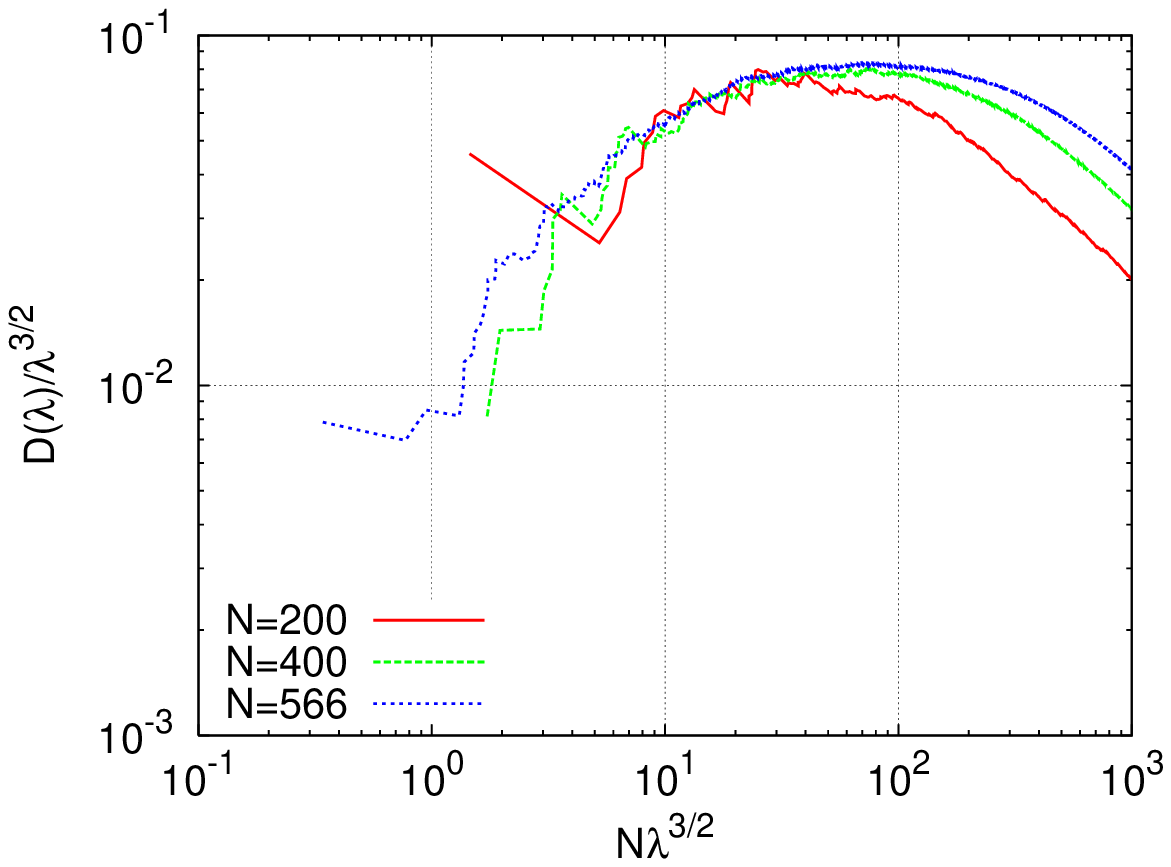}
\caption{Scaling plot of the averaged integrated density of states of the
Hessians for different system sizes.}
\label{hessfig}
\end{figure}

We have tested numerically the prediction $\chi_{\rm SG}\sim
N^{1/3}$ using the iteration procedure described in
Ref.~\cite{TAPiteration}. This method allows us to find many TAP
states even for large system sizes (see Ref.~\cite{TAPiteration}
for a discussion of the proximity of the iteration algorithm to a
dynamical critical point and the influence of this on the free energy
of the states found -- here we chose the proximity for each system
size in such a way that we get the same free energy range for all
system sizes).  Using systems with $N=100$, $200$, $283$, $400$, $566$
and $800$ at a temperature of $T = 1.0/\beta = 0.2$, we have
calculated the Hessian for every state found. The diagonalization of
the Hessians was done using arbitrary precision arithmetic (with an
accuracy of $50$ decimal digits) since these matrices are extremely
ill-conditioned and so they are hard to diagonalize: standard
packages, for instance the LAPACK routines, fail at the task.  The
eigenvalues were used to calculate $\chi_{\rm SG}$ as in
Eq.~(\ref{chiSG}). The results are shown in Fig.~\ref{chiSGfig}.

Surprisingly, $\chi_{\rm SG}$ appears to grow faster than
$N^{1/3}$. We believe, however, that this is a finite size effect.  To
back up this claim, we show in Fig.~\ref{hessfig} a scaling plot
of the averaged integrated eigenvalue density $D(\lambda)$ of the
Hessians of sizes $N=200, 400$ and $566$. 

The expectation is that in the thermodynamic limit this function goes
as $D(\lambda)=D_{\infty}(\lambda)\sim\lambda^{3/2}$ for small
$\lambda$ (corresponding to $\rho(\lambda)\sim\lambda^{1/2}$). For
finite $N$ there is a cutoff around $\lambda\approx N^{-2/3}$. The
natural expectation is that this cutoff is of the form
$D(\lambda)=D_{\infty}(\lambda)f(N\lambda^{3/2})$ with a scaling
function $f(x)$. This is verified in Fig.~\ref{hessfig}. The arguments
sketched above which lead to the prediction $\chi_{\rm SG}\sim
N^{1/3}$ for large $N$ can only be expected to be valid when the
interval in which $D(\lambda) \approx \lambda^{3/2}$ prevails is large enough. This
interval can be identified as the horizontal (or nearly horizontal)
stretch in Fig.~\ref{hessfig}.  Clearly, this interval is very small
as it is not even one decade for the available system sizes. The
conclusion is, therefore, that we cannot yet expect to see the
asymptotic scaling behaviour. It is not possible to go to larger system
sizes as the arbitrary precision diagonalization of the Hessians
becomes computationally too expensive.

\section{Free energy fluctuations and $J(p)$}
\label{appJp}

According to Ref.~\cite{Francesco}, the exact expression for the
quantity $J(p)$ for a finite number of replica symmetry breaking steps
$K$ and for the truncated model with Hamiltonian
\begin{equation}
\mathcal H = -\frac t2 \sum_{\alpha,\beta}q_{\alpha\beta}^2 - \frac w6
\sum_{\alpha,\beta,\gamma} q_{\alpha\beta}q_{\beta\gamma}q_{\gamma\alpha} -
\frac{y}{12}\sum_{\alpha,\beta}q_{\alpha\beta}^4
\end{equation}
is
\begin{equation}
J(p) = -\sum_{k,l=1}^{K+1} \mu_0(k)\mu_0(l)\log(p^2+\lambda(0;k,l))\;,
\label{Jexact}
\end{equation}
where
\begin{eqnarray}
\mu_r(k) &=& \left\{
\begin{array}{ll}
\frac{1}{p_k}-\frac{1}{p_{k-1}} & k>r-1 \\
\frac{1}{p_{r+1}} & k=r+1 \end{array}\right.\;\;,\\
p_k &=& \frac{2y}{w}q_K\frac{k-\frac 12}{K}\;,\\
\lambda(r;k,l) &=& 2y\frac{q_K^2}{K^2} \left(\frac 12 (k-1)^2+\frac 12 
(l-1)^2-r^2-\frac 16\right)\;,
\end{eqnarray}
and $q_K$ is the solution of
\begin{equation}
t-wq_K+y\left(1-\frac{1}{6K^2}\right)q_K^2 = 0\;.
\end{equation}
When the expressions for $\mu_r(k)$ and $\mu_r(l)$ are inserted, 
Eq.~(\ref{Jexact}) can be rewritten as
\begin{eqnarray}
J(p) &=& -2\sum_{k=1}^K \frac{1}{p_k} 
\log\frac{p^2+\lambda(0;k,K+1)}{p^2+\lambda(0;k+1,K+1)} - 
\log(p^2+\lambda(0;K+1,K+1))  \nonumber\\
&& -\sum_{k,l=1}^K \frac{1}{p_k 
p_l}\log\left(\frac{p^2+\lambda(0;k,l)}{p^2+\lambda(0;k+1,l)} 
\frac{p^2+\lambda(0;k+1,l+1)}{p^2+\lambda(0;k,l+1)}\right) .
\end{eqnarray}
We are interested in the behaviour for small $p$ where $J(p)$ diverges as $p$ 
approaches $\frac{yq_K^2}{3K^2}$. It is easy to see that the first two terms 
of the former expression
are well behaved in this limit. The divergence in $p$ must therefore come from 
the last term, which will be denoted by $\hat{J}(p)$. Defining
\begin{equation}
x^2 = \frac{K^2 p^2}{yq_K^2}-\frac 13
\end{equation}
and renumbering the sums to start from $0$ it can be cast in the form
\begin{eqnarray}
\hat{J}(p) &=& -\frac{w^2 (x^2+\frac 13)}{4y p^2}\sum_{k,l=0}^{K-1}
\frac{1}{\left(k+\frac 12\right)\left(l+\frac 12\right)} \nonumber\\
&&\times\log\left(\frac{x^2+k^2+l^2}{x^2+(k+1)^2+l^2}
\frac{x^2+(k+1)^2+(l+1)^2}{x^2+k^2+(l+1)^2}
\right)\;.
\label{Jhat}
\end{eqnarray}
While $p$ was a finite-dimensional wave vector in
Ref.~\cite{Francesco}, here we consider it as a proxy for the
self-energy as we are dealing with the SK model. Substituting
$\Sigma_R=cN^{-1/3}$ for $p^2$ and making the usual replacement
$K=c'tN^{1/6}$ (where the constant $c'$ is large enough to guarantee
stability) yields (to leading order in $N$) $x^2=\frac{{c'}^2 c
  t^2}{yq_\infty^2}-\frac 13$ and $\hat{J}=\delta
F^2=N^{1/3}f(t)$. The function $f(t)$ is defined by the remaining
prefactors and the sums (with upper bounds set to infinity) in
Eq.~(\ref{Jhat}). This shows that the typical sample-to-sample fluctuations 
of the free energy are of
order $N^{1/6}$.


\end{document}